\documentclass[aps,prd,twocolumn,amsfonts,amssymb,amsmath,showpacs,letterpaper,superscriptaddress,nofootinbib,altaffilletter,fontenc]{revtex4}

\usepackage[citecolor=blue,linkcolor=blue,colorlinks=true]{hyperref} 

\usepackage{color}
\usepackage{graphicx}
\usepackage{latexsym}
\usepackage{parskip}
\usepackage{float}
\usepackage{amsmath}
\usepackage{amssymb}
\usepackage{multirow}
\usepackage{ifthen}
\usepackage{slashed}
\usepackage{lineno}
\usepackage{makecell} 
\usepackage[mathscr]{euscript} 
\usepackage{mathrsfs} 
\usepackage{subfigure}
\usepackage{blindtext}
\usepackage{setspace}
\usepackage{stackrel}

\usepackage{multirow}
\usepackage[utf8]{inputenc}
\usepackage{comment}

\usepackage{tabularx,booktabs}
\newcolumntype{Y}{>{\centering\arraybackslash}X}
\usepackage{booktabs}
\usepackage{pifont}

\usepackage{array}

\newcolumntype{R}[1]{>{\RaggedRight}p{#1}}

\usepackage[normalem]{ulem}

\renewcommand{\today}{\number\day\space\ifcase\month\or
  January\or February\or March\or April\or May\or June\or
  July\or August\or September\or October\or November\or December\fi
  \space\number\year}

\def\be{\begin{equation}}
\def\ee{\end{equation}}
\def\bi{\begin{itemize}} 
\def\ei{\end{itemize}}
\def\ben{\begin{enumerate}}
\def\een{\end{enumerate}}

\def\apj{Astrophys. J.}
\def\apjl{Astrophys. J. Lett.}
\def\apjs{Astrophys. J. Supp. Ser. }

\def\physrep{Phys. Rep. }
\def\mnras{Mon. Not. Roy. Astron. Soc. }
\def\prl{Phys. Rev. Lett.}
\def\prd{Phys. Rev. D.}

\def\pasa{Proc. Astron. Soc. Pac.}

\begin{document}



\title{Characterizing the gravitational wave temporal evolution of the gmode fundamental resonant frequency for a core collapse supernova: A neural network approach}

\author{Alejandro~Casallas-Lagos} 
\affiliation{Department of Physics and Astronomy, Embry-Riddle Aeronautical University, Prescott, AZ 86301, USA}
\email[E-mail: ]{alejandroc.lagos@alumnos.udg.mx}
\affiliation{Departamento de Física, Universidad de Guadalajara, Guadalajara, Jal., 44430, M\'exico}

\author{Javier~M.~Antelis}
\email[E-mail: ]{mauricio.antelis@tec.mx}
\affiliation{Escuela de Ingeniería y Ciencias, Tecnológico de Monterrey, Monterrey, N.L., 64849, México}
\affiliation{Department of Physics and Astronomy, Embry-Riddle Aeronautical University, Prescott, AZ 86301, USA}

\author{Claudia~Moreno}
\affiliation{Department of Physics and Astronomy, Embry-Riddle Aeronautical University, Prescott, AZ 86301, USA}
\affiliation{Departamento de Física, Universidad de Guadalajara, Guadalajara, Jal., 44430, M\'exico}

\author{Michele~Zanolin}
\affiliation{Department of Physics and Astronomy, Embry-Riddle Aeronautical University, Prescott, AZ 86301, USA}

\author{Anthony~Mezzacappa}
\affiliation{Department of Physics and Astronomy, University of Tennessee, Knoxville, TN 37996-1200,
USA}

\author{Marek J. Szczepa\'nczyk} 
\affiliation{Department of Physics, University of~Florida, PO Box 118440, Gainesville, FL 32611-8440, USA}

\begin{abstract}
 We present a methodology based on the implementation of a fully connected neural network to estimate the gravitational wave (GW) temporal evolution of the gmode fundamental resonant frequency for a Core Collapse Supernova (CCSN). To perform the estimation, we construct a training data set, using synthetic waveforms, that serves to train the ML algorithm, and then use several CCSN waveforms to test the model. According to the results obtained from the implementation of our model, we provide numerical evidence to support the classification of progenitors according to their degree of rotation. The relative error associated with the estimate of the slope of the resonant frequency versus time for the GW from CCSN signals is within $13\%$ for the tested candidates included in this study. This method of classification does not require priors or templates, it is based on physical modelling, and can be combined with studies that classify the progenitor with other physical features. 
\end{abstract}


\maketitle

\section{Introduction}
\label{sec:Introduction}

\begin{figure*}[thpb]
    \centering
    \includegraphics[width=0.8\textwidth]{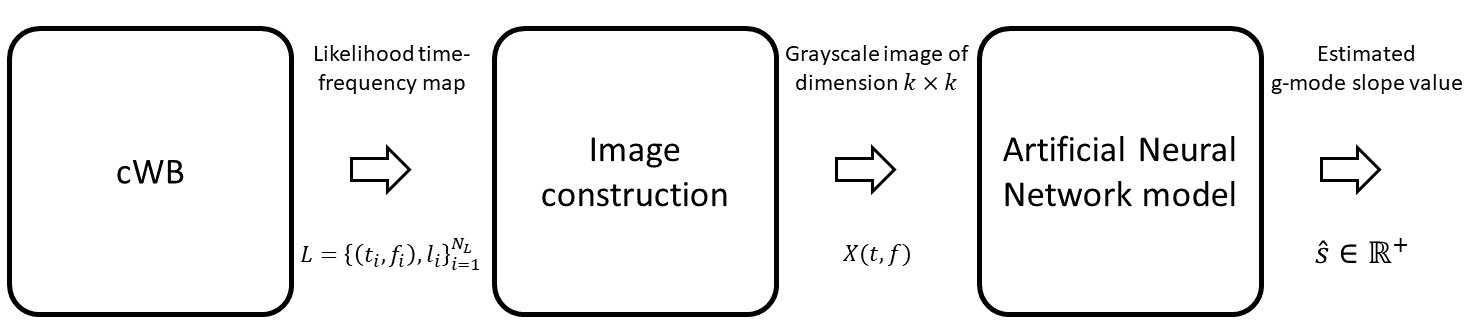}
\caption{
Graphical description of the follow-up deep learning method proposed to estimate physical parameters of GW from CCSN events detected with cWB -- specifically, the slope of the gmode oscillatory feature.
The cWB pipeline detects a GW event and provides reconstructed signal information, such as the likelihood time-frequency map $L$.
This time-frequency information is processed to construct a gray-scale image $X(t,f)$, which in turn is used as input for a deep learning regression model that estimates the value of the gmode slope $\hat{s}$.
}
    \label{fig:Methodology}
\end{figure*}

%
We are witnessing the era of ground-based gravitational wave (GW) detectors. Since 2015, the rate of confirmed events, the sensitivity and accuracy of the GW interferometers, and the detector network, have been improved to levels that open the door to new and complex Galactic sources of GW such as Core Collapse Supernovae (CCSNe). 
(For a review, see \cite{KoSaKa06,AbPaRa22}.)
A detection of this source type defines one of the main challenges in the near future for the Advanced LIGO~\cite{LIGOScientific:2014pky},  VIRGO~\cite{Acernese_2014}, and KAGRA~\cite{Aso_2013} detectors (LVK).
\\ \\
CCSNe designate the final life stage for a massive star $(M_{\odot}>8)$, a highly energetic process of stellar explosion recorded and observed since ancient times. (For a review, see \cite{JaMeSu16,Mueller16,MeEnMe20,Mueller20,BuVa21,Mezzacappa23}.)
The explosion process begins once the star's iron core mass exceeds its Chandrasekhar limit and collapses on itself. After core collapse, a compact, dense (above nuclear matter density $1-2\times 10^{14}g/cm^3$) 
star is created, a Proto-Neutron Star (PNS), whose physical properties are inherited from the progenitor star. Several processes involving different regions of the PNS are associated with the generation of high-frequency (above 100~Hz) GW: convective instabilities, convective overshoot, and accretion onto the PNS (e.g., see \cite{Andresen_2017,Mezzacappa_2020,MeMaLa23}).
\\ \\
A central problem in transient GW astronomy is to reconstruct the physical parameters associated with the source of the gravitational radiation when the signal is detected in laser interferometric data. This problem involves two parts: The identification of the relevant deterministic parameters, and the best procedure to estimate them. The GW from CCSN numerical simulations manifest as strongly stochastic signals
\cite{
Andresen_2017, 
Andresen:2018aom, 
Cerd_Dur_n_2013, 
HaKuKo15, 
HaKuKo18, 
HaKuNa16, 
KaKuTa18, 
KoIwOh11, 
KoIwOh09, 
KuFiTa22,
Kuroda_2017, 
KuKoTa16, 
Kuroda_2018, 
Mezzacappa_2020, 
MeMaLa23,
Morozova_2018, 
M_ller_2013, 
Murphy_2009, 
Nakamura_2022, 
O_Connor_2018, 
Ott_2013, 
Pajkos_2019,
PaWaCo21, 
Pan_2018, 
Pan_2021, 
Powell_2019, 
Powell_2020, 
Powell_2022, 
Radice_2019, 
RiZaAn23, 
Richers_2017, 
Scheidegger_2010, 
ShKuKo20, 
ShKuKo21, 
SrBaBr19, 
TaKo18, 
VaBu20, 
VaBuRa19, 
Warren_2020, 
YaMeMa15}; 
nevertheless, some features can be classified as deterministic. A feature that emerges from all CCSN numerical simulations is known as the 
\emph{gmode}
\cite{
Astone_2018, 
Bizouard_2021, 
Cerd_Dur_n_2013, 
KaKuTa18, 
KuKoTa16, 
Kuroda_2017, 
Mezzacappa_2020, 
MeMaLa23, 
Morozova_2018, 
M_ller_2013, 
Murphy_2009, 
O_Connor_2018, 
Pajkos_2019, 
Pan_2018, 
PaWaCo21, 
Radice_2019, 
ShKuKo20, 
ShKuKo21, 
SrBaBr19, 
Torres_Forn__2019, 
VaBuRa19, 
Warren_2020}.
\\ \\
This feature is recognizable in a time-frequency spectrogram as a continuous, strictly increasing, and to a first approximation linear feature, starting at around $100~Hz$ and increasing up to $\sim1-2~KHz$ with time after bounce. 
The rate of increase of the resonant frequency of the fundamental mode of vibration of the PNS (usually called fundamental gmode) is a deterministic feature believed to be strongly correlated with the degree of rotation of the progenitor and properties of the PNS.   The relationship between the slope of the gmode and progenitor rotation is discussed, for example, in 
\cite{
Pajkos_2019}. 
A first attempt to estimate the slope of the gmode with real interferometric noise was performed in
\cite{https://doi.org/10.48550/arxiv.2211.07878},
applying a chi-squared method to a low-order polynomial evolution of the resonant frequency. The authors applied the procedures on CCSN events identified by cWB, the flagship algorithm for the detection of GW bursts.  Other studies \cite{Bizouard_2021}, proposed an approach involving  normal mode decomposition, along with a polynomial interpolation and simulated Gaussian noise, to infer the time evolution of a combination of the mass and radius of the compact remnant.
\\ \\
In this work we use an optimized neural network approach for the estimation of the slope of the gmode from CCSN events detected by cWB. The results aim to quantify the capability to use the slope of the gmode to discriminate the degree of rotation of the progenitor, as well as other of its physical parameters. Accordingly, the numerical determination of the gmode slope constitutes a critical component of a framework for parameter estimation that can be used once the GW from a CCSN are detected by the LIGO,
Virgo, and KAGRA
detectors.
\\ \\
In the rest of the text we focus our attention on the estimation of the slope of the gmode. In order to estimate the gmode slope, we develop a deep neural network (DNN) model for regression. To do this, we use Coherent WaveBurst (cWB) \cite{Klimenko_2005, Klimenko_2016, Klimenko_2008, DRAGO2021100678}, a powerful computational pipeline designed to detect and reconstruct GW bursts with minimal assumptions about the morphology of the signal. We perform a simulation analysis in cWB on two different kinds of GW signals: The first is used to train our algorithm, providing known slopes associated with spectrograms that reveal a linear growth of the gmode. These signals will be designated as synthetic waveforms. Once the algorithm is trained from the synthetic waveforms 
\cite{Astone_2018}, 
a set of gravitational waveforms from CCSN simulations 
\cite{
Andresen_2017, 
Andresen:2018aom, 
Cerd_Dur_n_2013, 
Kuroda_2017, 
Mezzacappa_2020, 
Morozova_2018, 
O_Connor_2018} 
are included as testing data. Figure \ref{fig:Methodology} illustrates the different steps proposed in this manuscript.  
The accuracy of the DNN model is studied in order to determine the capability of the algorithm to estimate the slope of the gmode for an arbitrary CCSN signal in real interferometric data.   

The manuscript is organized as follows. 
%
Section~\ref{sec:methodology} describes the stages for the construction of the training and testing data sets, the cWB analyses performed to obtain the signal triggers and likelihood maps for the CCSN signals selected, the corresponding processing for the cWB outcomes, and the topology of the neural network model and the hyper-parameters that determine the performance of the neural network. In section~\ref{sec:results} we present the results of the implementation and the accuracy of the model. Finally, in section~\ref{sec:summary} we present the conclusions and future directions for this work.

\section{Methodology}
\label{sec:methodology}
In this section we describe the methodology carried out to assess the feasibility of estimating the slope of the gmode fundamental resonant frequency associated with CCSN GW events detected with cWB. 
The methodology uses ($\textbf{A}$) GW from CCSN signals, ($\textbf{B}$) cWB simulation analyses to obtain likelihood time-frequency maps of detected events, ($\textbf{C}$) processing of the likelihood time-frequency maps to construct an image, and ($\textbf{D}$) the DNN
model used to estimate the gmode slope value.

\begin{figure*}[thpb]
    \centering
    \includegraphics[width=0.35\textwidth]{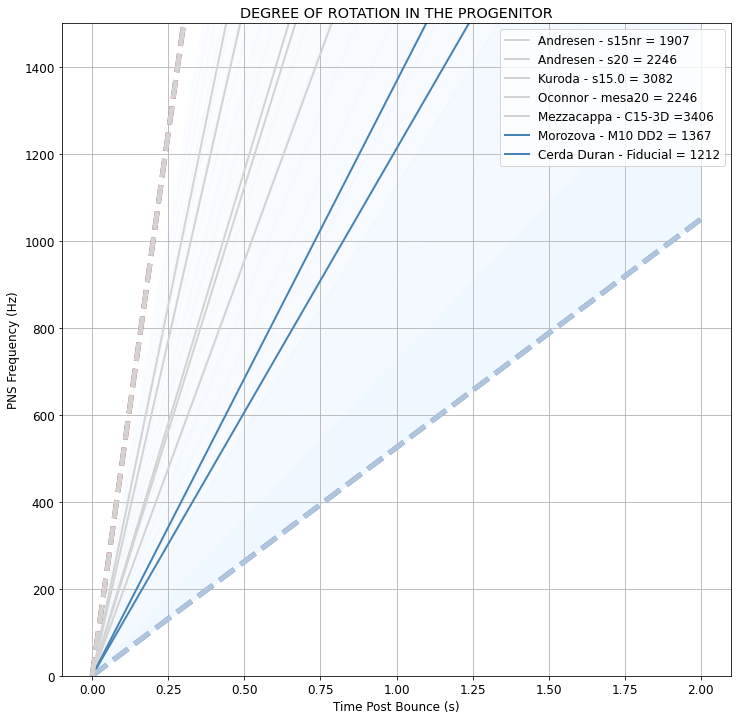}
    \includegraphics[width=0.35\textwidth]{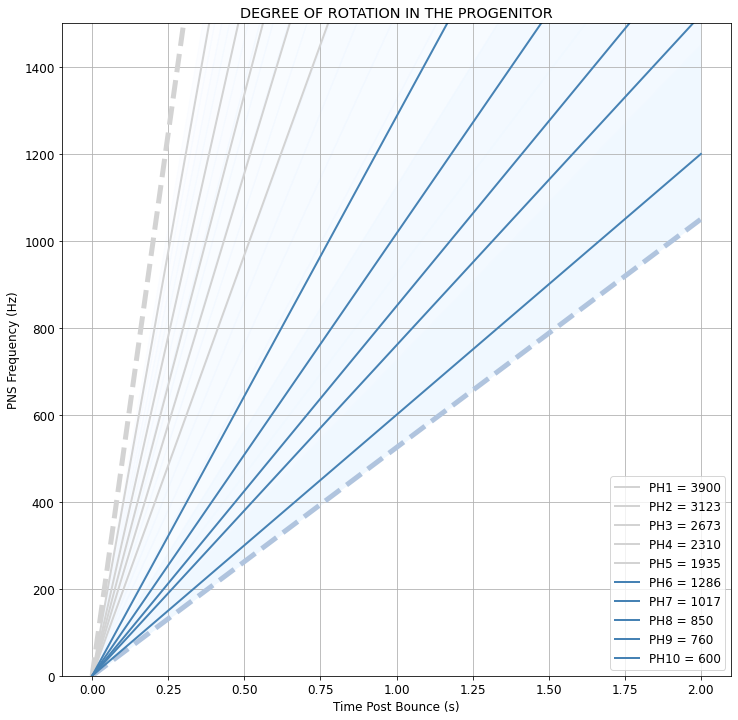}
    \caption{Range of gmode slopes associated with $(i)$ GW from CCSN signals included in this study [see section \ref{subsec:3Dsignals}] and  $(ii)$ 10 different synthetic signals [see section \ref{subsec:SyntheticSignals}].} 
    \label{fig:degree}
\end{figure*}

Figure \ref{fig:degree} illustrates the relationship between the gmode slopes of the ten synthetic models and the seven core collapse supernova models considered here. 
The lower and upper slope limits are defined by Equations (\ref{E:gmSR}) and (\ref{E:gmRR}), rspectively, and are obtained through the methodology described in this section. 
Less (more) inclined slopes are associated with rapidly (slowly) rotating progenitors.
\begin{equation}\label{E:gmSR}
    f_{SR}=525\;Hz/s.
\end{equation}
\begin{equation}\label{E:gmRR}
    f_{RR}=4990\;Hz/s.
\end{equation}

\subsection{GW from CCSN signals}

\subsubsection{Synthetic signals}
\label{subsec:SyntheticSignals}
\begin{figure*}[thpb]
 \centering
   \label{a}
    \includegraphics[width=0.31 \textwidth]{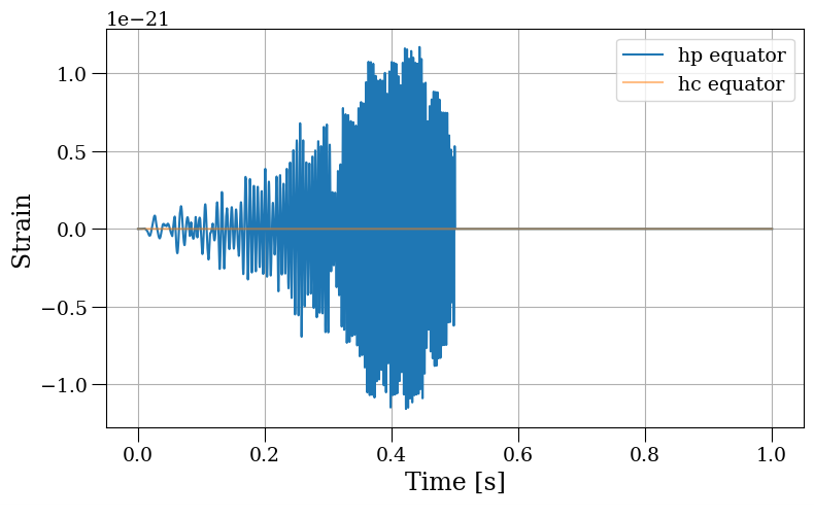}
   \label{b}
    \includegraphics[width=0.31 \textwidth]{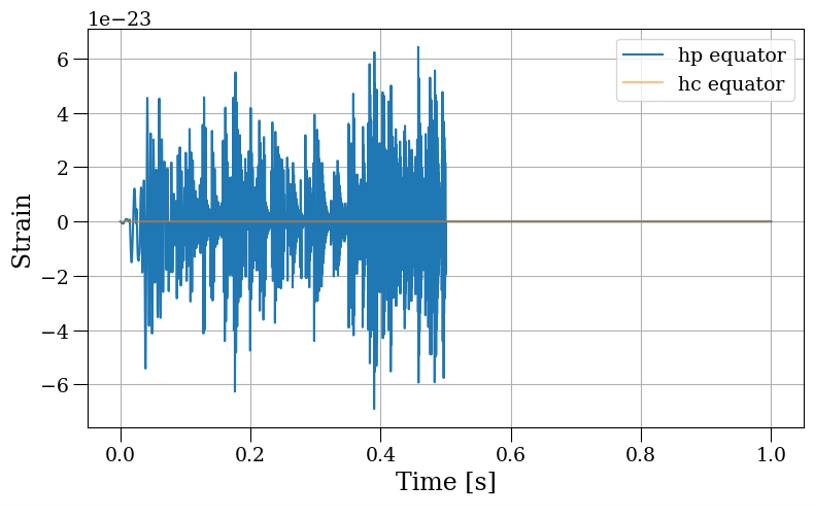}
   \label{c}
    \includegraphics[width=0.31 \textwidth]{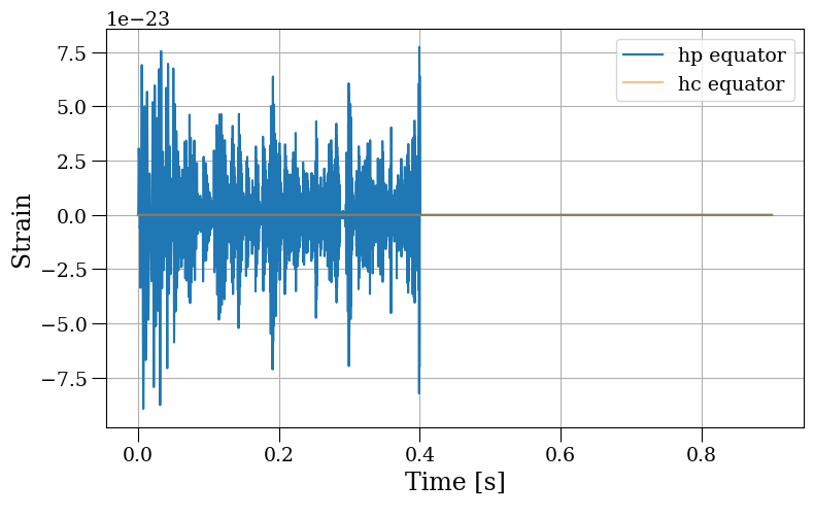}
   \label{a}
    \includegraphics[width=0.31 \textwidth]{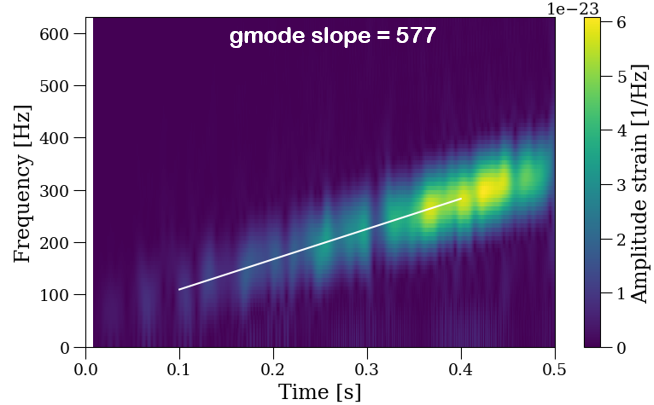}
   \label{b}
    \includegraphics[width=0.31 \textwidth]{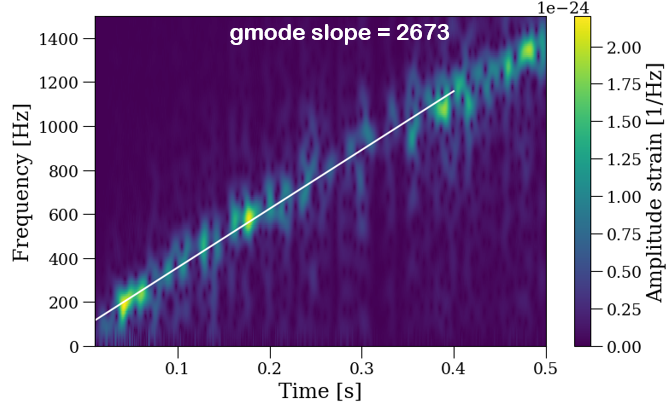}
   \label{c}
    \includegraphics[width=0.31 \textwidth]{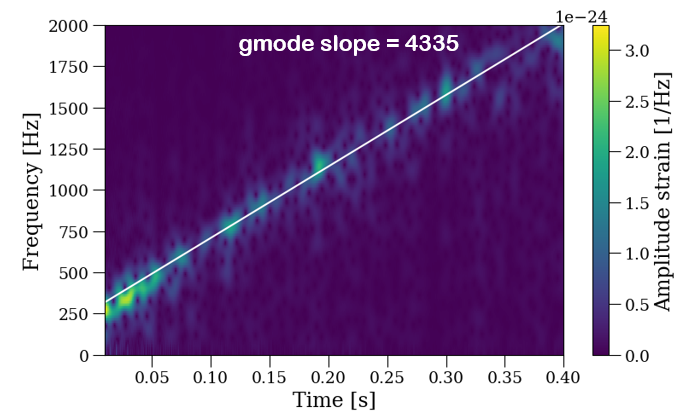}  
\caption{
Example of three synthetic CCSN GW signals with gmode slopes of 577, 2673 and 4335 Hz/s representing rapidly, moderately, and slowly rotating progenitors, respectively. The top panel shows the strain signals, while the bottom panel shows their spectrograms. The solid white line in the spectrograms traces the rising frequency over time.
}
   \label{fig:ExampleOfSyntheticGW}
\end{figure*}
We created stochastic signals with increasing frequency over time, observable in their spectrograms, thus emulating CCSN GW signals containing the gmode feature.
The synthetic signals are to be used in cWB simulation analyses to obtain a training data set of likelihood time-frequency maps with known values of the gmode slope. 
These signals were created based on the damped harmonic oscillator with an external stochastic driving force; i.e., a second order, non-homogeneous differential equation, as proposed in recent work (see Equation (1) in~\cite{Cerd_Dur_n_2013}). The solution to such differential equations is performed numerically, and the choice of several parameters such as the duration, and initial and final frequencies (which encode the gmode) can be modified easily to obtain different solution signals.
Even though these synthetic signals do not carry any physical information, they are highly beneficial because it is straightforward to vary their associated parameters and because the computational cost to generate them is very low.
Therefore, we can obtain signals that resemble GW from CCSN, with the gmode feature, and for each signal we directly have the value of the gmode slope, $s$, which is simply computed as the difference between the higher and lower frequency divided by its duration.
We generated 100 different synthetic CCSN GW signals with gmode slopes ranging from 500 to 5000~Hz/s, to cover the full range of expected slopes from rapidly to slowly rotating progenitors
reported in the literature.
%
Figure \ref{fig:ExampleOfSyntheticGW} shows a sample of three synthetic CCSN GW signals included in this study, with gmode slope values of $577$, $2673$, and $4335$ Hz/s.
\subsubsection{Numerical simulation signals}
\label{subsec:3Dsignals}

\begin{figure*}[thpb]
 \centering
   \label{a}
    \includegraphics[width=0.9 \textwidth]{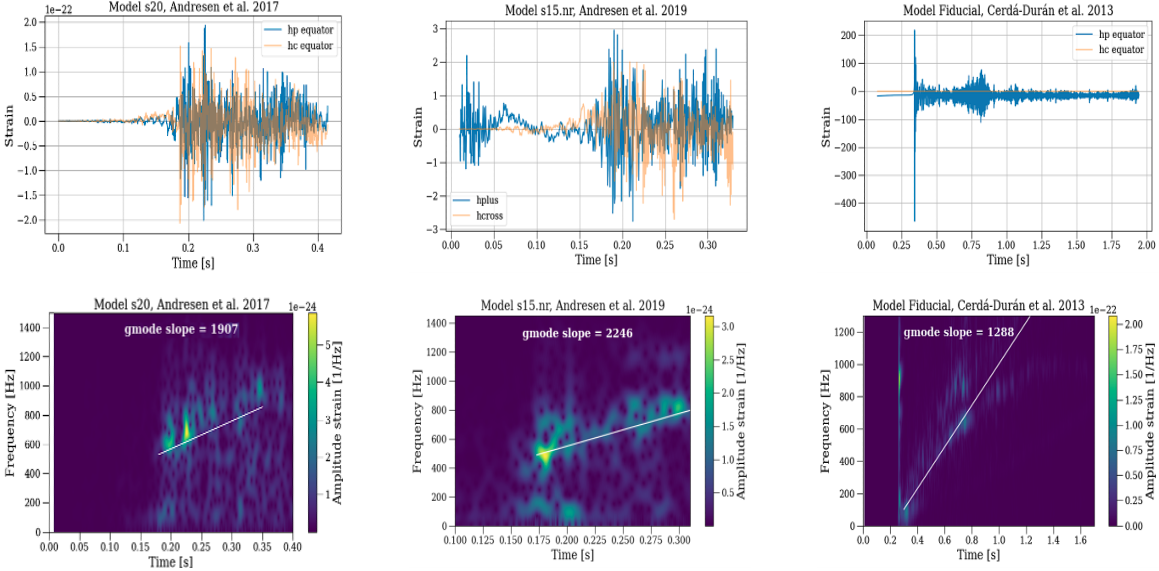}
   \label{b}
    \includegraphics[width=0.9 \textwidth]{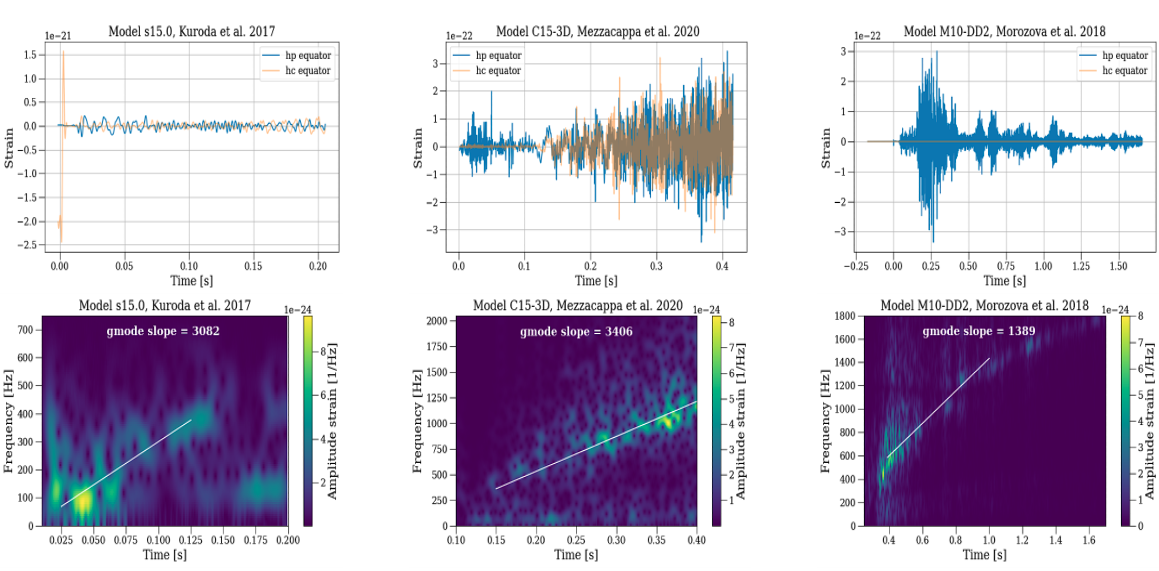}
   \label{c}
    \includegraphics[width=0.6 \textwidth]{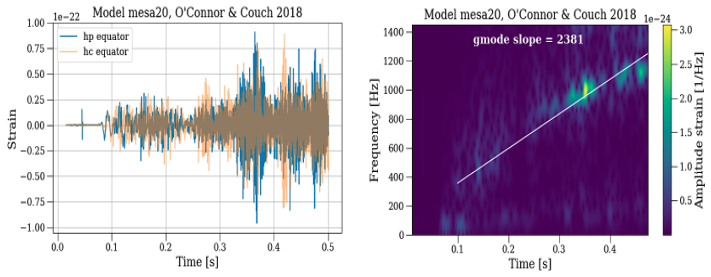}
\caption{
 Strain signals (left panels) and time-frequency evolution spectrograms (right panels) for CCSN GW models:  
 Model \textit{s20} from Andresen et al. 2017, 
 Model \textit{s15.nr} from Andresen et al. 2019,
 Model \textit{Fiducial} from Cerdá-Durán et al. 2013, 
 Model \textit{s15.0} from Kuroda et al. 2017, 
 Model \textit{C15-3D} from Mezzacappa et al. 2020, 
 Model \textit{M10-DD2} from Morozova et al. 2018, 
 and Model \textit{mesa20} from O'Connor and Couch 2018.
 The solid white line in the spectrograms traces the gmode.
 }
\label{fig:ExampleOf3DGW}
\end{figure*}

We also used CCSN GW signals from 2D and 3D numerical simulations, all of which contain the gmode feature, associated with slowly (including non-rotating), moderately, and rapidly rotating progenitors, in order to cover the different simulation scenarios 
reported 
in the literature.
%
This set of GW from CCSN signals are used in our cWB simulation analyses as a test data set, with their known values of the gmode slope in the likelihood time-frequency maps.
Specifically, we selected the following set of GW from CCSN signals computed from different progenitors and degree of rotation:
%
%
%
\begin{itemize}
    \item Model \textit{s20} from Andresen et al. 2017 \cite{2019MNRAS.486.2238A}. The GW signal is extracted from a 3D approximately general relativistic radiation (neutrino) hydrodynamics simulation with a  20~$M_{\odot}$ non-rotating progenitor. 
    \item Model \textit{s15.nr} from Andresen et al. 2019 \cite{Andresen_2017}. The GW signals are extracted from three different models based on 3D approximately general relativistic radiation (neutrino) hydrodynamics simulations with a single progenitor with a zero-age main-sequence mass (ZAMS) of 15~$M_{\odot}$, solar metallicity, and with different rotation rates $0\;rad/s$, $0.2\;rad/s$, and $0.5\;rad/s$ 
    \item Model \textit{Fiducial} from Cerdá-Durán et al. 2013 \cite{Cerd_Dur_n_2013}. This is a GW signal generated from a 2D general relativistic hydrodynamics simulation associated with a low-metallicity, rapidly-rotating progenitor, with a zero-age main-sequence mass of $35M_{\odot}$ whose initial central angular velocity is $2\; rad/s$.
    \item Model \textit{s15.0} from Kuroda et al. 2017 \cite{Kuroda_2017}. For this signal, the GW emission is obtained from a 3D general relativistic radiation (neutrino) hydrodynamics with a $14\;M_{\odot}$, solar metallicity, non-rotating progenitor. 
    \item Model \textit{C15-3D} from Mezzacappa et al. 2020 \cite{Mezzacappa_2020}. For this signal the GW emission is computed for a 3D approximately general relativistic radiation (neutrino) hydrodynamics simulation with a  non-rotating 15~$M_{\odot}$ progenitor of solar metallicity.
     \item Model \textit{M10-DD2} from Morozova et al. 2018 \cite{Morozova_2018}. This signal was generated from a 2D approximately general relativistic radiation (neutrino) hydrodynamics CCSN simulation with a 10~$M_{\odot}$ progenitor with solar metallicity and moderate rotation:  0.2~rad/s.
     \item Model \textit{mesa20} from O'Connor and Couch 2018 \cite{O_Connor_2018}. In this model, the GW emission is modelled from a 3D approximately general relativistic radiation (neutrino) hydrodynamics CCSN simulation with a
    20~$M_{\odot}$, solar metallicity, non-rotating progenitor. 
    %
    %
\end{itemize}
Note that all of these CCSN GW signals are from 3D numerical simulations except for the signals models \textit{Fiducial} and \textit{M10-DD2}, which correspond to 2D simulations.
In addition, these signals were used in recent studies involving targeted searches \cite{Szczepa_czyk_2021}, false detection rates \cite{Slutsky_2010}, and sensitivity analyses of GW's from CCSNe \cite{Cavagli__2020}, using strain data of the LIGO, VIRGO, and KAGRA detectors.
Figure \ref{fig:ExampleOf3DGW} shows the strain signals and the spectrograms of signal models \textit{s20}, \textit{s15.nr}, \textit{Fiducial}, \textit{s15.0}, \textit{C15-3D}, \textit{M10-DD2}, and \textit{mesa20}. Note how the spectrograms manifest the gmode feature.
%

\subsection{cWB simulation analyses}
Coherent WaveBurst (cWB) is a standard method for detecting and reconstructing GW embedded in strain data acquired with the 
LIGO,
VIRGO,
and KAGRA
~\cite{Aso_2013} 
detectors.
The method uses minimal assumptions about the signal morphology~\cite{Klimenko_2005, Klimenko_2008, Klimenko_2016, DRAGO2021100678}, which is a necessary condition in the search for un-modelled GW's, as those from CCSNe.
The cWB algorithm (1) searches for coincident signal power across detectors by projecting the multi-detector data onto the wavelet (i.e., time-frequency) domain using the Wilson-Daubechiers-Meyer transform~\cite{Necula_2012}, (2) identifies a collection of coherent time-frequency components with amplitudes above noise levels, and (3) clusters them to obtain a likelihood time-frequency map $L = \{ (t_i,f_i) ,l_i \}_{i=1}^{N_L}$, where $l_i$ is the likelihood point value at time $t_i$ and frequency $f_i$, and $N_L$ is the number of time-frequency points. Figure \ref{fig:LikelihoodTimeFrequencyMap}  shows the likelihood time-frequency map $L$ for a detected event from a synthetic GW signal.
We use in this study the likelihood time-frequency map $L$ to estimate the gmode slope value because it contains the significant time-frequency information that is used to reconstruct the detected GW signal. cWB simulation analyses were performed using LIGO data from the second half of the third observing run (O3b) with a two-detector network (L1 and H1).
The aim was to obtain distributions of likelihood time-frequency maps of detected GW from CCSNe, to train and to test the deep learning algorithm that estimates the value of the gmode slope.
In these cWB analyses, known GW from CCSN signals were injected every 50~s, at a distance of $1~kpc$ and with equatorial orientation, into the detector strain data.
Then, the search for GW is carried out, and for each detected event, the likelihood time-frequency map $L$ is computed, along with some reconstructed signal attributes.
All cWB simulation analyses were performed in two separate stretches of strain data.
The first, comprising 1~day of coincident data, was used to obtain the training data, while the second stretch of data, comprising 8~days of coincident data, was used to obtain the testing data. 
In addition, our cWB analyses were performed separately with the synthetic and with the CCSN GW signals containing the characteristic gmode feature, as was described below.
The set of synthetic signals constructed as part of this study (see subsection \ref{subsec:SyntheticSignals}) was used in the first stretch of strain data, to obtain the training data set, $\mathcal{D}_{train} = \{ L_j , s_j \}_{j=1}^{N_{train}}$, where $L_j$ and $s_j$ are the likelihood time-frequency map and the gmode slope value of the $j$-th detected event, respectively, and $N_{train}$ is the number of training instances.
Furthermore, the set of GW from CCSN signals (see subsection \ref{subsec:3Dsignals}) was used in the second stretch of strain data, to obtain the test data set of likelihood time-frequency maps, $\mathcal{D}_{test} = \{ L_j , s_j \}_{j=1}^{N_{test}}$, where $N_{test}$ is the number of test instances.
It is important to remark at this point that training and test data sets are mutually exclusive, which is a necessary condition to assess the robustness of the machine learning algorithm used to perform the estimation of the gmode slope with unknown GW from CCSN signals.

\subsection{Image construction}
\label{subsec: image construction}
\begin{figure}
    \centering
    \includegraphics[width=0.21\textwidth]{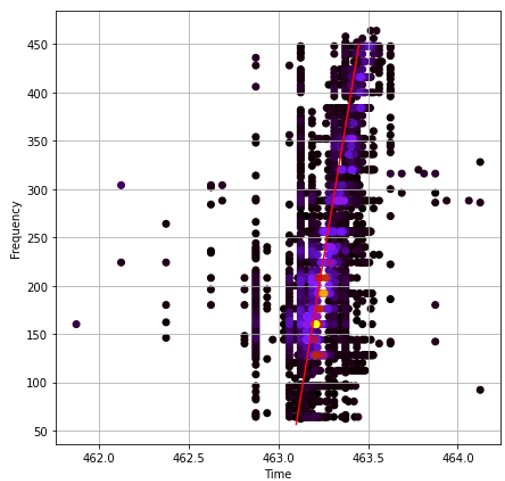}
    \includegraphics[width=0.20\textwidth]{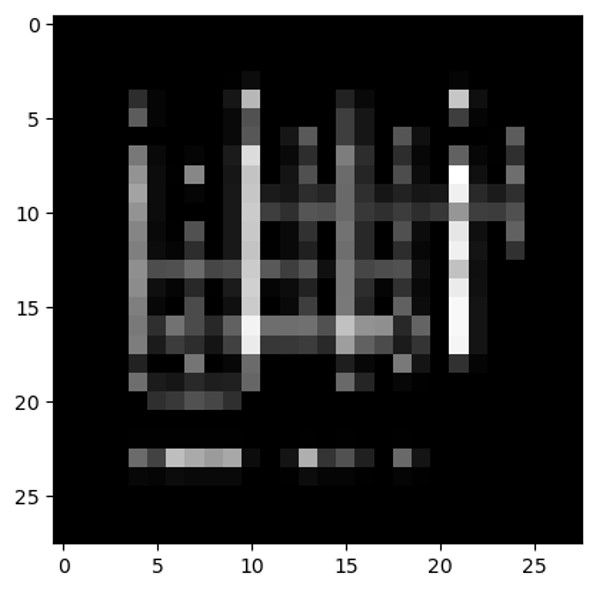}
    \caption{
    Likelihood time-frequency map $L$ of a detected event from a synthetic signal and its corresponding two-dimensional data matrix $X(t,f)$ of dimension $k \times k$ with $k=28$.
    }
    \label{fig:LikelihoodTimeFrequencyMap}
\end{figure}
The likelihood time-frequency map, $L = \{ (t_i,f_i) ,l_i \}_{i=1}^{N_L}$, varies across detected events in the number of points, the frequency range, and the time range.
Therefore, it is necessary to produce a data representation with standard dimensions so that it can be used as input to the machine learning model that estimate the slope of the gmode.
Hence, the goal is to define a function $f(\cdot)$ that maps $L$ into a two-dimensional data matrix $X(t,f)$; that is, $f \colon L \to X(t,f)$, where the width ($t$-dimension) and height ($f$-dimension) are the same for any detected GW. Given $L = \{ (t_i,f_i) ,l_i \}_{i=1}^{N_L}$, the image construction is carried out through the following procedure.
%
First, we select the time-frequency point that has the maximum likelihood value, $\{ t_{m}, f_{m} \}$.
%
%
Then, we select a region around $t_{m}$ in the interval $[t_{m}-\delta_t, \ t_{m}+\delta_t]$ (width in the $t$-dimension of 2$\cdot \delta_t$), and around $f_{m}$ in the interval 
$[50~Hz, 750~Hz]$, where the cWB reconstructed signal is more accurate. The length of $\delta_t$ is fixed at $0.3~s$ such that the time interval is large enough to contain the primary evolution of the early high-frequency gmode present in the GW from the CCSN models considered in this study [see section \ref{subsec:3Dsignals}].
%
%
This region is then transformed into a high-resolution, gray-scale image where the intensity is given by the likelihood value of the corresponding time-frequency points, while pixels with no corresponding time-frequency points are set to zero.
%
Finally, this gray-scale image is downsized to dimension $N_{r} \times N_{c}$ to obtain the final two-dimensional data matrix or image $X(t,f)$, where $N_{r}$ and $N_{c}$ represent the number of rows and columns, respectively.
%
%
Equivalently, the data matrix or image $X(t,f)$ can be flattened to construct the $N$-dimensional column vector $\textbf{x} \in \mathbb{R}^{N}$, where $N = N_{r} \cdot N_{c}$.
Figure \ref{fig:LikelihoodTimeFrequencyMap} shows the two-dimensional data matrix $X(t,f)$, with dimension $N_{r} = N_{c} = 28$, for a likelihood time-frequency map, $L$, of a detected event given a synthetic signal.
\subsection{Deep Neural Network model for regression}
To estimate the gmode slope of CCSN GW events detected with cWB, we use a deep learning regression model 
The input to this model is the column vector representation $\textbf{x} \in \mathbb{R}^{N}$ of the processed likelihood time-frequency map, $X(t,f)$, while the output is the estimated gmode slope value $\hat{s}$.
Note that the gmode slope values are continuous, real, and positive (i.e., $s \in \mathbb{R}^{+}$), ranging from 500~Hz/s (lower limit for rapidly-rotating progenitors) to 5000~Hz/s (upper limit for slowly-rotating progenitors).
Hence, there are several regression methods that can be used to address this task; for instance, linear and polynomial models, decision trees, and artificial neural networks~\cite{Cuoco_2020, George_2018, Antelis_2022, Chan_2020, Mukherjee:2017, Cavagli__2020, L_pez_2021, Morales2020}.
For this study, we selected a fully-connected deep neural network (DNN) regression model because it can learn both linear and non-linear relationships between the input and output data, it is more appropriate for handling large-dimensional input data, and it offers high performance at a low computational cost.

%
%
%
%
DNN are machine learning models inspired by biological neural network models of the brain, consisting of many interconnected processing units known as neurons, which vaguely mimic biological neurons~\cite{George_2018, Antelis_2022, Bishop2006}.
The structure of a DNN comprises an input layer, one or more hidden layers, and an output layer, thus resembling a brain neural network~\cite{Antelis_2022, Morales2020, Mukherjee:2017}.
The input layer consists of nodes that receive the input data and pass them directly into the first hidden layer for further processing, whereas hidden and output layers consist of many neurons~\cite{Antelis_2022, Bishop2006} interconnected by weighted synaptic links.
%
%
In a DNN, the information flows from the input towards the output while being processed in the layers through the following function:
\begin{equation}
    f:\mathbb{R}^{m}\longmapsto \mathbb{R}^{n} ,
\end{equation}
where $m$ and $n$ denote the number of nodes or neurons of two successive layers.
Therefore, the $j$-th neuron in a layer (hidden or output) with $n$ neurons is connected to all of the $m$ outputs of the preceding layer, via the weighted synaptic connections \cite{Bishop2006}, in such a way that the neuron produces the output $y_{j}$ as a function of linear combinations of the input information as follows:
\begin{equation}
    y_{j} = g \left( \sum_{i=1}^{m} w_{i,j} x_{i} \right),
\end{equation}
where $w_{i,j}$ are the weights connecting all $m$ inputs to the $j$-th neuron and $g(\cdot)$ is a bounded, differentiable, real, and nonlinear function known as an activation function \cite{George_2018, L_pez_2021}.
The activation function allows the nonlinearity between the input--output relationship. 
In regression problems, typical activation functions are linear, sigmoid, or the rectified linear unit \cite{Bishop2006}.
Note that the number of nodes of the input layer corresponds to the input variables (in our case, $N$ inputs), the number of neurons in the output layer corresponds to the output variables (in our case, one output), whereas the number of hidden layers, the number of neurons in each layer, and the activation functions are hyper-parameters that can be freely varied to obtain different DNN models.
Therefore, the synaptic weights are the parameters that are fitted from a training data set. We considered five DNN models with different numbers of hidden layers and neurons.
Table \ref{tab:DNNmodels} presents the technical details of the DNN models, which are named M1, M2, M3, M4, and M5.
In all models, the activation functions correspond to the rectified linear unit (ReLu) in the hidden layers and to the linear function in the output layer.
The training of the models (i.e., the fitting of the synaptic weights) was based on the back-propagation learning algorithm using the root-mean-squared propagator (MNSprop) with the mean-squared error as a loss function, a learning rate of 0.001, a batch size of 512 samples of the training data, and 300 epochs.
\begin{table}
\centering
\caption{Architectural description of the deep neural network (DNN) regression models used to assess the estimation of the gmode slope from likelihood time-frequency maps.
}
\begin{tabular}{lccc} 
	\hline
	\hline
	\multicolumn{1}{c}{   } & \multicolumn{1}{c}{\# of hidden} & \multicolumn{1}{c}{\# of   } & \multicolumn{1}{c}{{ \# of  }} \\
    \multicolumn{1}{c}{   } & \multicolumn{1}{c}{layers      } & \multicolumn{1}{c}{neurons } & \multicolumn{1}{c}{{weights }} \\
	\hline
	\hline
	  M1  & 1 & 16               & 12,577  \\
	  M2  & 2 & 32-16            & 25,665  \\
	  M3  & 3 & 64-32-16         & 52,865 \\
	  M4  & 4 & 128-64-32-16     & 111,361  \\
	  M5  & 5 & $\;\;\;$ 256-128-64-32-16 $\;\;\;$ & 244,737  \\
	\hline
	\hline
\end{tabular}
\label{tab:DNNmodels}
\end{table}
%
\section{Results}
\label{sec:results}
%
In this section we present the results of two analyses devoted to assessing the performance of the estimation of the gmode slope of CCSN GW  events using DNN regression models.
\subsection{Hold-out cross-validation with $\mathcal{D}_{train}$}
\label{subsec:study1}
\begin{figure}
    \centering
    \includegraphics[width=0.5\textwidth]{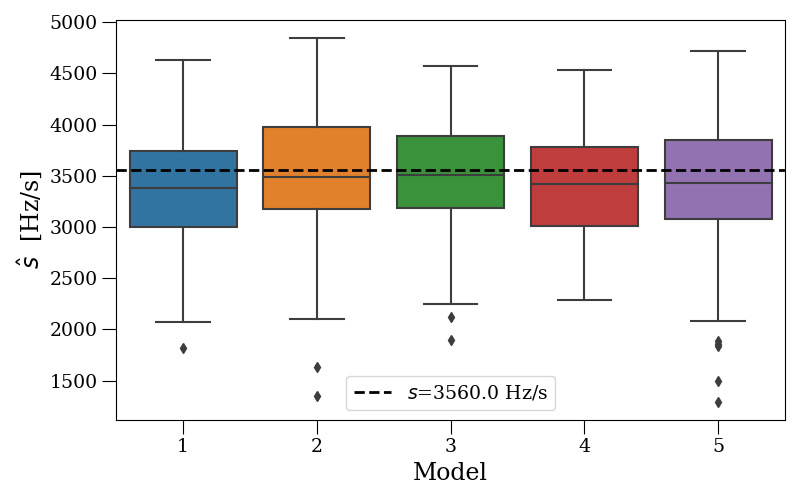}
    \includegraphics[width=0.5\textwidth]{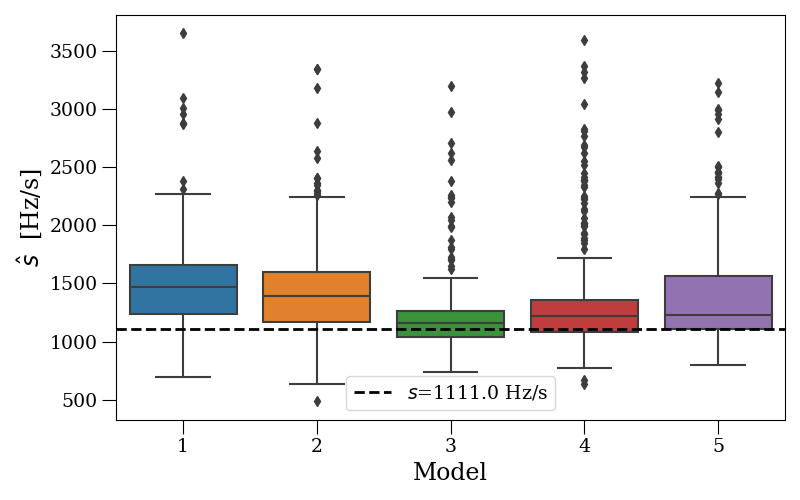}
    \caption{
    Distribution of estimated values $\hat{s}$ achieved with the five DNN models in the hold-out cross-validation procedure with the data set $\mathcal{D}_{train}$ for the specific case of real values of $s=3560.0$ Hz/s (upper panel) and $s=1111.0$ Hz/s (bottom panel).
    }
    \label{fig:study1_examples}
\end{figure}

We first assessed the effectiveness and reliability of the proposed DNN model for the estimation of the gmode slope, testing the training data set of likelihood time-frequency maps, $\mathcal{D}_{train}$, through a hold-out cross-validation (HOCV) procedure \cite{Bishop2006, Burrows_2021}, where the entire data set was randomly split into two parts for training (70$\%$) and for testing (30$\%$).
This procedure was repeated 30 times to account for the randomness of the process and to be able to compute distributions of the performance metrics.
Note that in each repetition the training and testing data are mutually exclusive.
The training set is used to fit the weights of the DNN model, while the test set is used to asses the model performance \cite{Antelis_2022, Morales2020}.
To assess the performance, we used the following metrics: $(i)$ the coefficient of determination ($r^2$), which measures for every model the linear correlation between the known slopes $(s)$ present in the likelihood maps and the corresponding estimated slopes $(\hat{s})$; $(ii)$ the root-mean-square error (rmse), computed as the square root of the difference between the estimated $\hat{s}$ and known values $s$ of the gmode slope, which serves to discriminate
how far from the mean the estimated slopes are, and finally $(iii)$ the mean-absolute-percentage error (mape), to evaluate the precision of each model performing the estimation of the gmode slope. These performance metrics provide support to evaluate the accuracy of each model from different perspectives and clarify the outputs obtained. Table \ref{tab:Study1_Metrics} shows the average values of the performance metrics achieved with the five DNN models.
\\ \\
According to the scores presented in 
the table, we conclude that model three (M3) (see Table \ref{tab:DNNmodels}) exhibits the best performance in estimating the slope of the gmode, among the five different DNN architectures (M1 to M5), because (1) it has the higher linear correlation $(0.76)$ expressed through the $r^2$ coefficient and (2) the lowest residuals $(594.64)$ and percentage error $(21\%)$ reported by the rmse and mape, respectively. 
\begin{figure}
    \centering
    \includegraphics[width=0.5\textwidth]{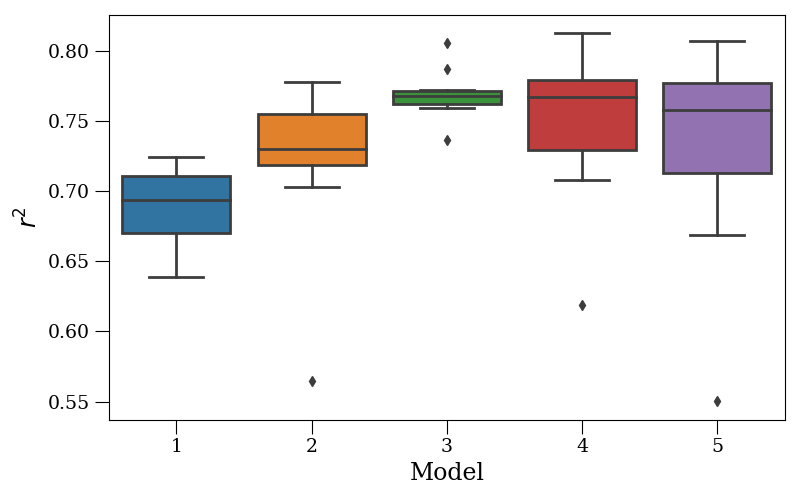}
    \includegraphics[width=0.5\textwidth]{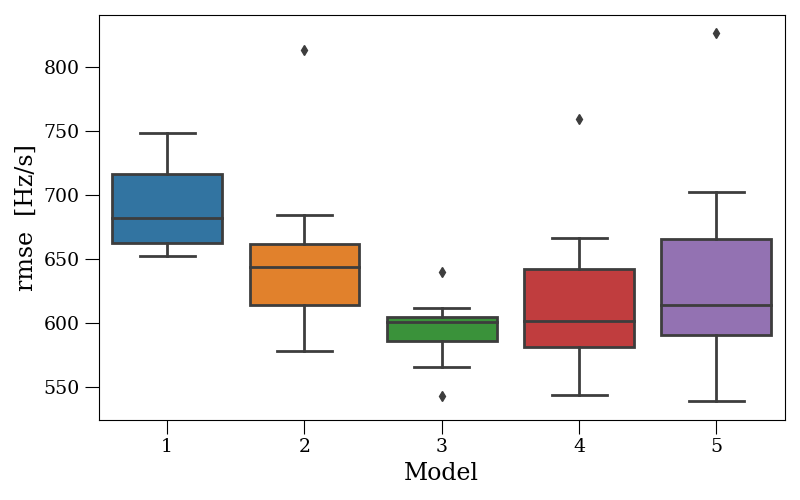}
    \includegraphics[width=0.5\textwidth]{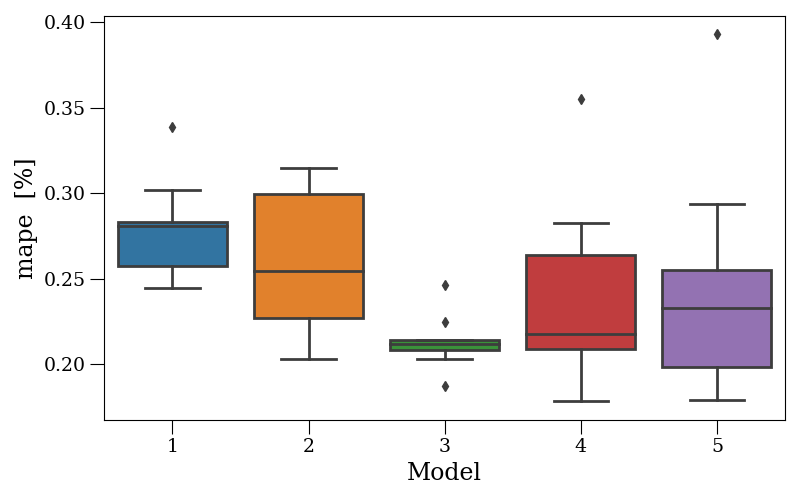}
    \caption{Distribution of the performance metrics $r^2$, rmse and mape achieved with the five DNN regression models in the hold-out cross-validation procedure with the data set $\mathcal{D}_{train}$.}
    \label{fig:study1}
\end{figure}
Figure \ref{fig:study1} illustrates the performance metrics  
for each model. Model 3 (in green) shows a lower dispersion with respect to the mean of the estimated slopes for the gmode, and lower residuals compared with those associated with the remaining models: M1 (blue), M2 (orange), M4 (red) and M5 (purple).         
\begin{table*}
\centering
\caption{Summary (mean $\pm$ standard deviation) of the performance metrics $r^2$, rmse, mae, and mape achieved with the DNN regression models in the hold-out cross-validation procedure with the data set $\mathcal{D}_{train}$.}
\begin{tabular}{c| c@{\hspace*{1em}} c@{\hspace*{1em}} c@{\hspace*{1em}} c@{\hspace*{1em}} c@{\hspace*{1em}} c@{\hspace*{1em}}}
	\hline
	\hline
    Metric & M1 & M2 & M3 & M4 & M5 \\
	\hline
	\hline
    $r^2$       & $0.68\pm 0.02$       & $0.72\pm 0.05$       & \textbf{0.76$\pm$ 0.01}       & $0.74\pm 0.05$       & $0.73\pm 0.07$       \\
    rmse [Hz/s] & $690.57\pm 32.86$ \, & $651.27\pm 65.08$ \, & \textbf{594.64$\pm$ 26.31} \, & $616.79\pm 62.03$ \, & $636.85\pm 81.94$ \, \\
    mape [\%]   & $0.27\pm 0.02$       & $0.25\pm 0.04$       & \textbf{0.21$\pm$ 0.01}       & $0.23\pm 0.05$         & $0.24\pm 0.06$         \\
	\hline
	\hline
\end{tabular}
\label{tab:Study1_Metrics}
\end{table*}
To give a more individual characterization of the different DNN models, Figure \ref{fig:study1_examples} illustrates how distinct architectures estimate a single slope contained in the training data set. This figure clarifies the fact that estimation performed by model M3 produces the best fit compared with the other DNN architectures.     

\subsection{Estimation of the gmode slope of 3D CCSN GW signals}
\label{subsec:study 3}
\begin{figure*}
    \centering
    \includegraphics[width=0.48\textwidth]{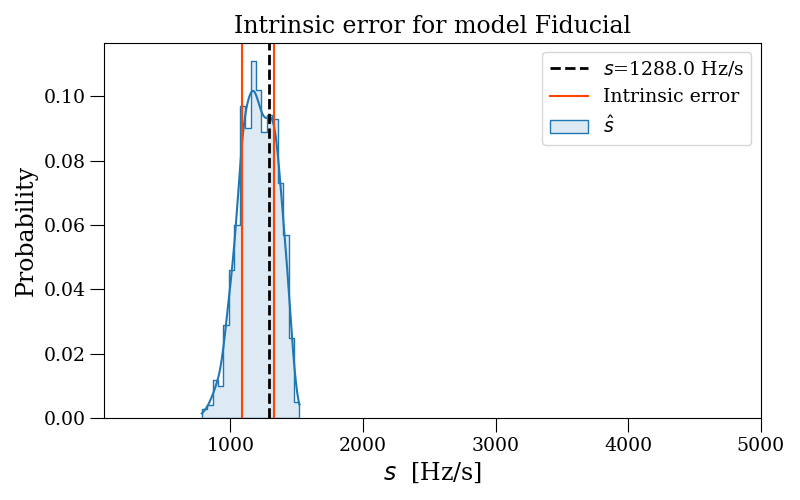}
    \includegraphics[width=0.48\textwidth]{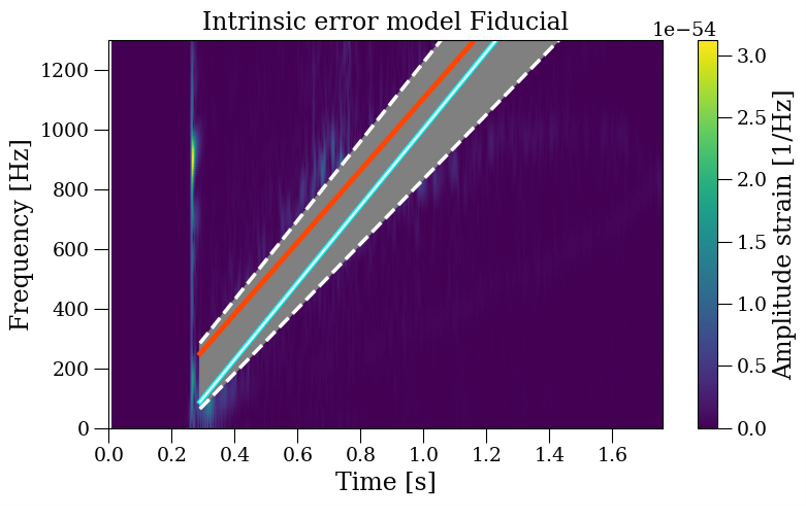}
    \includegraphics[width=0.48\textwidth]  {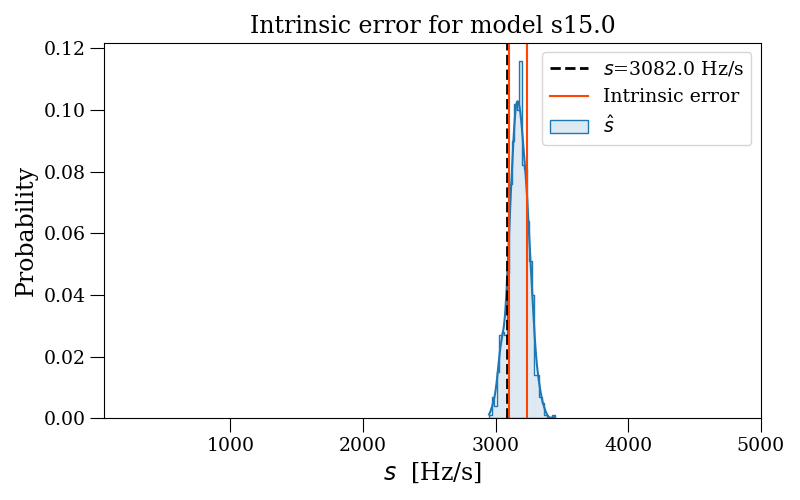}
    \includegraphics[width=0.48\textwidth]{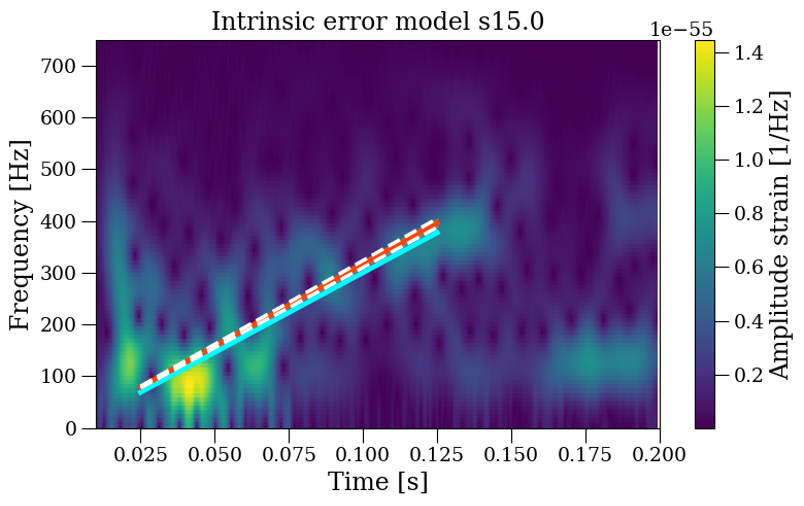}
    \includegraphics[width=0.48\textwidth]{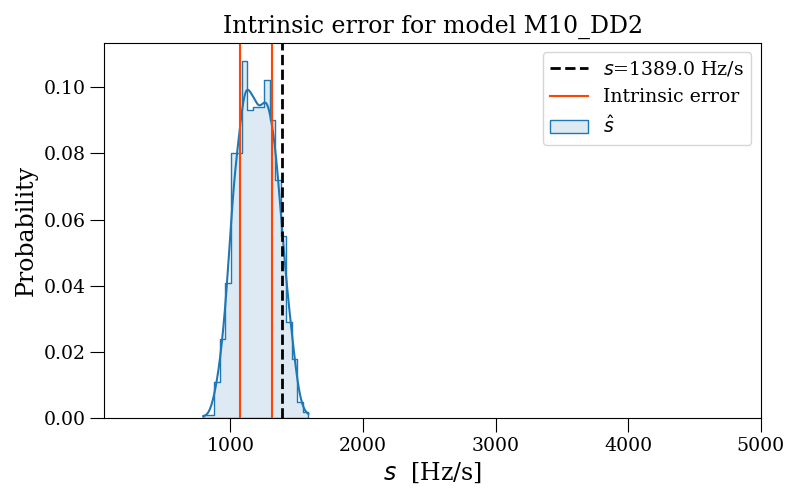}
    \includegraphics[width=0.48\textwidth]{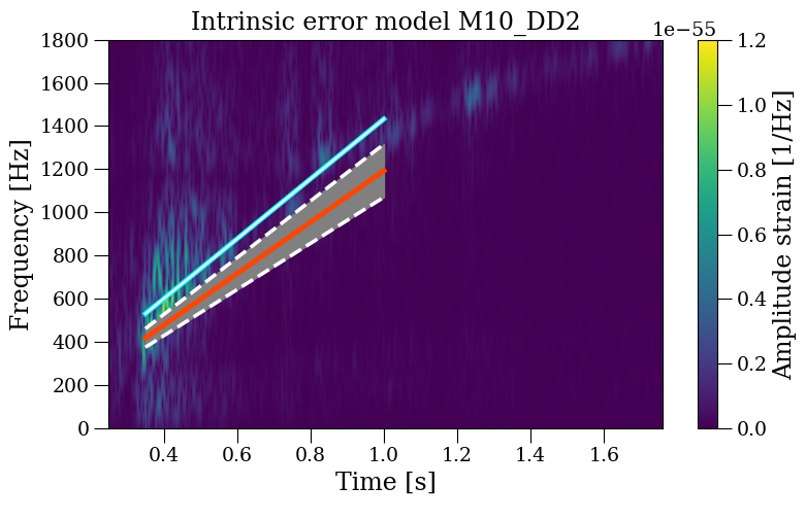}
    \includegraphics[width=0.48\textwidth]{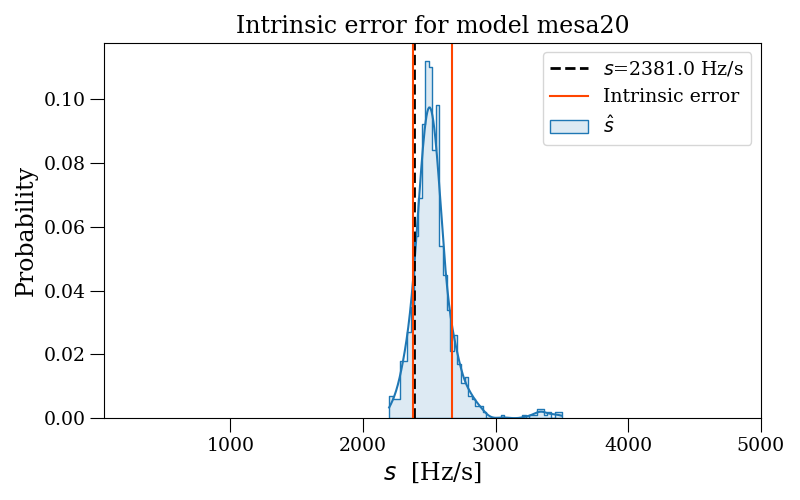}
    \includegraphics[width=0.48\textwidth]{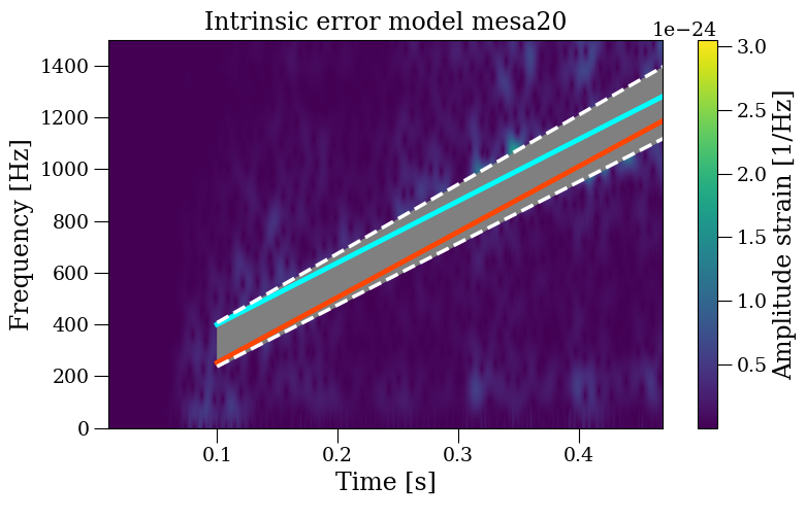}
    \caption{Estimation of slopes and intrinsic error associated to  different CCSN signals included in this study. Left column shows the estimated slopes (blue) and the intrinsic error associated to the estimation approach (red solid lines). At the right column the representation of the intrinsic error on the spectrogram of the GW from CCSN signal is illustrated in white dashed lines. In magenta the slope of the gmode, in red the mean of the estimated slopes obtained using the DNN.}
    \label{fig:results1}
\end{figure*}

\begin{figure*}
    \centering
    \includegraphics[width=0.48\textwidth]{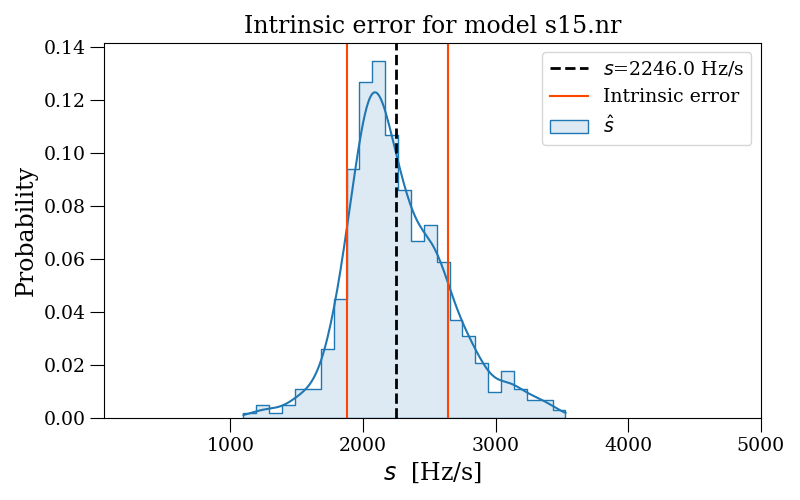}
    \includegraphics[width=0.48\textwidth]{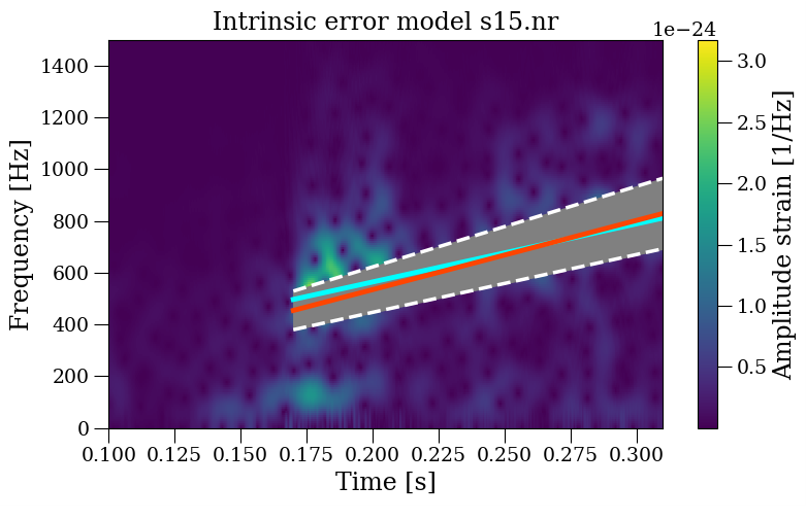}
    \includegraphics[width=0.48\textwidth]{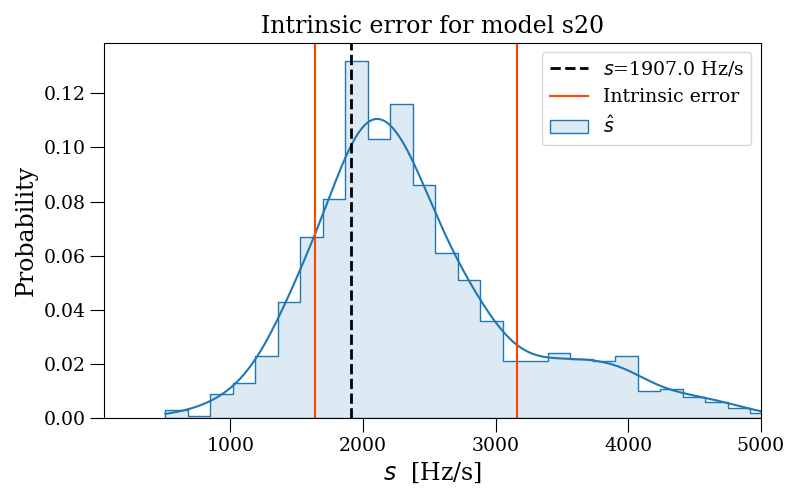}
    \includegraphics[width=0.48\textwidth]{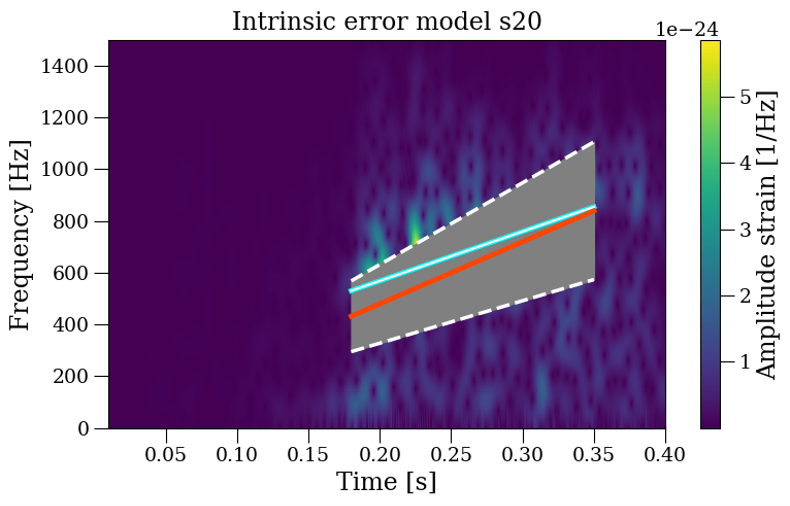}
    \includegraphics[width=0.48\textwidth]{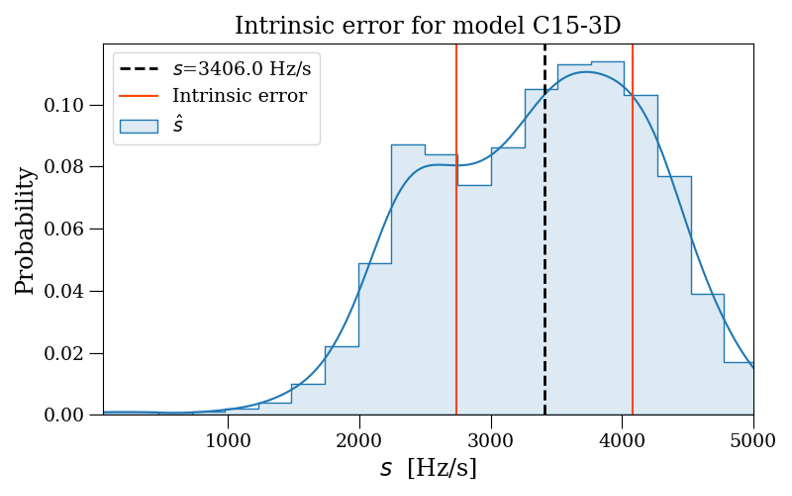}
    \includegraphics[width=0.48\textwidth]{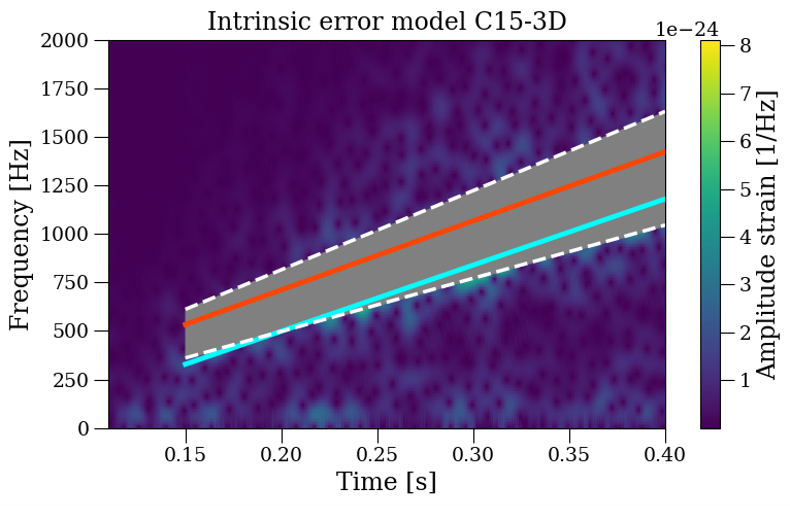}
    \caption{Estimation of slopes and intrinsic error associated with different CCSN signals included in this study. The left column shows the estimated slopes (blue) and the intrinsic error associated with the estimation approach (red solid lines). In the right column, the intrinsic error is represented on the spectrogram 
    with white dashed lines. Magenta designates the slope of the gmode, and red designates the mean of the estimated slopes obtained using the DNN.}
    \label{fig:results2}
\end{figure*}

After the design [Section \ref{sec:methodology}], construction [Section \ref{subsec: image construction}], and successive sanity checks performed on the training data set, $\mathcal{D}_{train}$ [Section\ref{subsec:study1}], we use the M3-DNN architecture [See table \ref{tab:DNNmodels}], 
along with the processed likelihood maps for the GW from CCSN signals, as a test data set, $\mathcal{D}_{test}$, in order to perform the estimation of the gmode slope.
\begin{table*}
\centering
\caption{gmode slope estimation results. Each column contains the values associated with the slope, $s$, estimated slope, $\hat{s}$, and its corresponding standard deviation, RMSE and MAPE. All of this is for the M3-DNN architecture.}
\begin{tabular}{c| c@{\hspace*{1em}} c@{\hspace*{1em}} c@{\hspace*{1em}} c@{\hspace*{1em}} c@{\hspace*{1em}} c@{\hspace*{1em}}}
	\hline
	\hline
    Model & $s~[Hz/s]$ & $\hat{s}~[Hz/s]$ & RMSE $[Hz/s]$ & MAPE \\
	\hline
	\hline
    \textit{Fiducial} from Cerdá-Durán et al. 2013 & $1288$       & \textbf{1204.78$\pm$ 120.75}       & $142.37$       & $0.09$       \\
     \textit{s15.0} from Kuroda et al. 2017 \, & $3082$ & \textbf{3169.36$\pm$ 67.52} \, & $108.68$ \, & $0.03$ \, \\
    \textit{M10-DD2} from Morozova et al. 2018 & $1389$       & \textbf{1193.24$\pm$ 122.76}       & $228.54$         & $0.14$         \\
    \textit{mesa20} from O'Connor and Couch 2018 & $2381$       & \textbf{2525.21$\pm$ 146.58}       & $204.95$         & $0.06$         \\
    \textit{s15.nr} from Andresen et al. 2019 & $2246$       & \textbf{2263.17$\pm$ 380.58}       & $380.59$         & $0.13$         \\
    \textit{s20} from Andresen et al. 2019 & $1907$       & \textbf{2399.42$\pm$ 759.80}       & $904.59$         & $0.33$         \\
    \textit{C15-3D} from Mezzacappa et al. 2020 & $3406$       & \textbf{3358.91$\pm$ 619.03}       & $719.98$         & $0.20$         \\
	\hline
	\hline
\end{tabular}
\label{tab:Results}
\end{table*}
$\mathcal{D}_{test}$ is only composed of processed likelihood maps obtained from CCSN signals [see \ref{subsec:3Dsignals}] that were not considered in the training process; therefore, the estimation of the gmode slope is carried out on $\mathcal{D}_{test}$, an unknown set of signals for the DNN architecture. Table \ref{tab:Results} contains the results of the implementation for the DNN architecture [see Table \ref{tab:DNNmodels}] on the different GW from CCSN signals included in this study [see section \ref{sec:methodology}]. The results reflect a mean standard deviation in the estimation of the gmode slope of $266.33~Hz/s$ and a mean-absolute-percentage error (mape) for this sample of $13\%$, which include GW from CCSN signals for different progenitor (ZAMS) masses, metallicities, and rotation rates.   
In Figures \ref{fig:results1} and \ref{fig:results2}, we show the distribution of estimated slopes (in blue) for each CCSN GW signal included in $\mathcal{D}_{test}$ (left column). The vertical, black, dashed line stands for the value of the slope, while the red solid lines define the intrinsic error associated with the estimation process, meaning that the range of slopes belonging to this interval covers the values of the estimation. In the right column, the spectrogram of each CCSN GW signal  illustrates the intrinsic error depicted in the histogram in the left column. Solid magenta lines indicate the value of the slope for every model, solid red lines indicate the mean of the estimated slopes obtained from the DNN architecture, while dashed white lines indicate the intrinsic estimation error. 
An estimation of the GW temporal evolution of the gmode fundamental
resonant frequency for a core collapse supernova was recently presented in \cite{https://doi.org/10.48550/arxiv.2211.07878}, using a chi squared approach  
in the context of a multimessenger analysis for the identification and parameter estimation of the Standing Accretion Shock Instability (SASI) with neutrino and GW signals. Following our methodology based on the implementation of a neural network (see sections \ref{sec:methodology} and \ref{subsec:study 3}), the estimation of the slope of the gmode fundamental resonant frequency for a CCSN is improved by $85\%$ for model \textit{s15.0} from Kuroda et al. 2017 when compared with the value presented in the study of the SASI. This fact reveals that the implementation of a neural network model exhibits a substantial improvement when compared with chi-squared-based estimation. 
Finally, to frame our results, we also consider the variation of the gmode slope with progenitor mass and EOS. Table \ref{tab:prog_mass} lists the gmode slopes of the models considered here, by descending order in progenitor mass. If we consider two models with the same rotation and EOS -- e.g., Model s20 and Model C15-3D -- we see that the variation of the slope is significant, with Model s20 having a slope of 1907 Hz/s and Model C15-3D having a much larger slope of 3406. Of course, there is more than just the progenitor dependence yielding different results here. There are model dependencies [approximations of the physics, different numerical methods deployed, different simulation codes, different grid resolutions adopted, differences in the input physics used (e.g., the weak interaction physics), etc.], though the UT--ORNL and MPA models are very similar in most respects. Nonetheless, the progenitor mass dependence of the gmode slope is large, which will necessitate multimessenger signals in order to break the redundancy of the dependence of the gmode slope on both the progenitor rotation and mass if we are to use the gmode slope to cull information about the progenitor's rotation. The EOS dependence is less significant. For example, comparing Models s15.0 and C15-3D, which are both initiated from the same progenitor mass and are both nonrotating, the slopes are 3082 Hz/s and 3406 Hz/s, respectively. Again, here too some of the difference can be attributed to model dependencies. Model s15.0 is general relativistic, with a minimum set of neutrino weak interactions, whereas Model C15-3D is only approximately general relativistic, but deploys an extensive weak interaction set. Moreover, while the progenitor mass may be the same in these two cases, the progenitor structure is different \cite{WoWe95,WoHe07}. Nonetheless, the gmode slope seems to be much more sensitive to the progenitor mass.


\begin{table*}
\centering
\caption{Variation of the gmode slope across progenitor mass and EOS.}
\begin{tabular}{c| c@{\hspace*{1em}} c@{\hspace*{1em}} c@{\hspace*{1em}} c@{\hspace*{1em}} c@{\hspace*{1em}} c@{\hspace*{1em}}}
	\hline
	\hline
    Model & $s~[Hz/s]$ & $\hat{s}~[Hz/s]$ & EOS & Mass $M_{\odot}$ & Rotation \\
	\hline
	\hline
    \textit{Fiducial}, Cerdá-Durán et al. 2013 & $1288$ & \textbf{1204.78} & LS220 & 
    $35$ & 2~rad/s \\
    \textit{mesa20}, O'Connor and Couch 2018 & $2381$ & \textbf{2525.21}    & SFHo & $20$ & -          \\
    \textit{s20}, Andresen et al. 2017 & $1907$       & \textbf{2399.42}  & LS220 & $20$ & -        \\
    \textit{s15.0}, Kuroda et al. 2017 & $3082$ & \textbf{3169.36} & SFHx & $15$ & - \\
    \textit{C15-3D}, Mezzacappa et al. 2020 & $3406$ & \textbf{3358.91} & LS220& $15$ & -\\ 
    \textit{s15.nr}, Andresen et al. 2019 & $2246$  & \textbf{2263.17} & LS220 & $15$ & 0.5~rad/s        \\ 
    \textit{M10-DD2}, Morozova et al. 2018 & $1389$       & \textbf{1193.24} & LS220 & $10$ & 0.2~rad/s\\
	\hline
	\hline
\end{tabular}
\label{tab:prog_mass}
\end{table*}
\section{Summary}
\label{sec:summary}
We incorporate a set of synthetic CCSN GW signals (see section \ref{sec:methodology}) to train a DNN model (section \ref{subsec:study1}) to estimate the slope associated with the gravitational wave temporal evolution of the gmode fundamental resonant frequency present in CCSN GW signals (section \ref{subsec:3Dsignals}). We quantified the accuracy of distinct DNN architectures using three different performance metrics to evaluate the accuracy of every model under different topologies, as presented in table \ref{tab:Study1_Metrics}, searching for the more accurate configuration to achieve estimation the gmode slope. 
Our analysis and implementation of such DNN architecture indicates, we can estimate the slope of the gmode fundamental resonant frequency in real interferometric data with an accuracy of $82\%$ within $2.3~kpc$, $65\%$ within $3.0~kpc$, and $52\%$ within $10~kpc$, for galactic sources and different idiosyncrasies of GW's from CCSNe (section \ref{subsec:3Dsignals}). Tables \ref{tab:galactic1} and \ref{tab:galactic2} shows the gmode slope estimation with its corresponding STD for seven different galactic sources; $1.0~kpc$, $2.3~kpc$, $3.1~kpc$, $4.3~kpc$, $5.4~kpc$, $7.3~kpc$ and $10~kpc$ respectively.  
The results obtained using our methodology reflect that, this implementation could be applied to develop parameter estimation in upcoming LIGO scientific runs. We leave this aspect for future publications.  
\begin{table*}
\centering
\caption{Variation of the gmode estimated slope across different galactic distances 1.0~kpc, 2.3~kpc, 3.1~kpc and 4.3~kpc.}
\begin{tabular}{c| c@{\hspace*{1em}} c@{\hspace*{1em}} c@{\hspace*{1em}} c@{\hspace*{1em}} c@{\hspace*{1em}}
c@{\hspace*{1em}}}
	\hline
	\hline
    Model & $s~[Hz/s]$ & 1.0~kpc & 2.3~kpc & 3.1~kpc & 4.3~kpc\\
	\hline
	\hline
    \textit{Fiducial}, Cerdá-Durán et al. 2013 & $1288$ & 1204.78 &  1345$\pm$ 221 & 1575 $\pm$ 496 & 1954 $\pm$ 596\\
    \textit{mesa20}, O'Connor and Couch 2018 & $2381$ & 2525.21 & 2689 $\pm$ 145 & 2903 $\pm$ 312 & 3312 $\pm$ 396\\
    \textit{s20}, Andresen et al. 2017 & $1907$  & 2399.10 & 2614$\pm$ 223 & 3012 $\pm$ 496 & 3324 $\pm$ 342\\
    \textit{s15.0}, Kuroda et al. 2017 & $3082$ & 3169.36 & 3243$\pm$ 109 & 3472 $\pm$ 221 & 3723 $\pm$ 443\\
    \textit{C15-3D}, Mezzacappa et al. 2020 & $3406$ & 3358.91 & 3443 $\pm$ 237 & 3743 $\pm$ 396 & 3978 $\pm$ 234\\
    \textit{s15.nr}, Andresen et al. 2019 & $2246$  & 2263.17 & 2509 $\pm$ 207 & 2689 $\pm$ 441 & 2945 $\pm$ 504\\
    \textit{M10-DD2}, Morozova et al. 2018 & $1389$ & 1193.24 & 1349 $\pm$ 115 & 1576 $\pm$ 396 & 1608 $\pm$ 503\\
	\hline
	\hline
\end{tabular}
\label{tab:galactic1}
\end{table*}
\begin{table*}
\centering
\caption{Variation of the gmode estimated slope across different galactic distances 5,4~kpc, 7.3~kpc and 10~kpc.}
\begin{tabular}{c| c@{\hspace*{1em}} c@{\hspace*{1em}} c@{\hspace*{1em}} c@{\hspace*{1em}}}
	\hline
	\hline
    Model & 5.4~kpc & 7.3~kpc & 10~kpc\\
	\hline
	\hline
    \textit{Fiducial}, Cerdá-Durán et al. 2013 & 2172 $\pm$ 598 & 2560 $\pm$ 698 & 3012 $\pm$ 723\\
    \textit{mesa20}, O'Connor and Couch 2018 & 3576 $\pm$ 696 & 3976 $\pm$ 621 & 4217 $\pm$ 876\\
    \textit{s20}, Andresen et al. 2017 & 3508 $\pm$ 554 & 4295 $\pm$ 662 & 4796 $\pm$ 883\\
    \textit{s15.0}, Kuroda et al. 2017 & 3998 $\pm$ 554 & 4209 $\pm$ 754 & 4873 $\pm$ 952\\
    \textit{C15-3D}, Mezzacappa et al. 2020 & 4110 $\pm$ 512 & 4675 $\pm$ 370 & 4975 $\pm$ 876\\ 
    \textit{s15.nr}, Andresen et al. 2019 & 3309 $\pm$ 555 & 3775 $\pm$ 875 & 4175 $\pm$ 576\\
    \textit{M10-DD2}, Morozova et al. 2018 & 1934 $\pm$ 634 & 2375 $\pm$ 772 & 2775 $\pm$ 902\\
	\hline
	\hline
\end{tabular}
\label{tab:galactic2}
\end{table*}
\\
\section*{Acknowledgement}
This work was supported by CONACyT Network Project No. 376127 {\it Sombras, lentes y ondas gravitatorias generadas por objetos compactos astrofísicos}.
ACL acknowledges a CONACYT scholarship. 
CM wants to thank PROSNI-UDG for support.
AM was supported by the National Science Foundation Gravitational Physics Theory Program (PHY 1806692 and 2110177).
%

\bibliography{paper}

\begin{thebibliography}{75}%
\makeatletter
\providecommand \@ifxundefined [1]{%
 \@ifx{#1\undefined}
}%
\providecommand \@ifnum [1]{%
 \ifnum #1\expandafter \@firstoftwo
 \else \expandafter \@secondoftwo
 \fi
}%
\providecommand \@ifx [1]{%
 \ifx #1\expandafter \@firstoftwo
 \else \expandafter \@secondoftwo
 \fi
}%
\providecommand \natexlab [1]{#1}%
\providecommand \enquote  [1]{``#1''}%
\providecommand \bibnamefont  [1]{#1}%
\providecommand \bibfnamefont [1]{#1}%
\providecommand \citenamefont [1]{#1}%
\providecommand \href@noop [0]{\@secondoftwo}%
\providecommand \href [0]{\begingroup \@sanitize@url \@href}%
\providecommand \@href[1]{\@@startlink{#1}\@@href}%
\providecommand \@@href[1]{\endgroup#1\@@endlink}%
\providecommand \@sanitize@url [0]{\catcode `\\12\catcode `\$12\catcode
  `\&12\catcode `\#12\catcode `\^12\catcode `\_12\catcode `\%12\relax}%
\providecommand \@@startlink[1]{}%
\providecommand \@@endlink[0]{}%
\providecommand \url  [0]{\begingroup\@sanitize@url \@url }%
\providecommand \@url [1]{\endgroup\@href {#1}{\urlprefix }}%
\providecommand \urlprefix  [0]{URL }%
\providecommand \Eprint [0]{\href }%
\providecommand \doibase [0]{http://dx.doi.org/}%
\providecommand \selectlanguage [0]{\@gobble}%
\providecommand \bibinfo  [0]{\@secondoftwo}%
\providecommand \bibfield  [0]{\@secondoftwo}%
\providecommand \translation [1]{[#1]}%
\providecommand \BibitemOpen [0]{}%
\providecommand \bibitemStop [0]{}%
\providecommand \bibitemNoStop [0]{.\EOS\space}%
\providecommand \EOS [0]{\spacefactor3000\relax}%
\providecommand \BibitemShut  [1]{\csname bibitem#1\endcsname}%
\let\auto@bib@innerbib\@empty
\bibitem [{\citenamefont {{Kotake}}\ \emph {et~al.}(2006)\citenamefont
  {{Kotake}}, \citenamefont {{Sato}},\ and\ \citenamefont
  {{Takahashi}}}]{KoSaKa06}%
  \BibitemOpen
  \bibfield  {author} {\bibinfo {author} {\bibfnamefont {K.}~\bibnamefont
  {{Kotake}}}, \bibinfo {author} {\bibfnamefont {K.}~\bibnamefont {{Sato}}}, \
  and\ \bibinfo {author} {\bibfnamefont {K.}~\bibnamefont {{Takahashi}}},\
  }\href {\doibase 10.1088/0034-4885/69/4/R03} {\bibfield  {journal} {\bibinfo
  {journal} {Reports on Progress in Physics}\ }\textbf {\bibinfo {volume}
  {69}},\ \bibinfo {pages} {971} (\bibinfo {year} {2006})},\ \Eprint
  {http://arxiv.org/abs/astro-ph/0509456} {arXiv:astro-ph/0509456 [astro-ph]}
  \BibitemShut {NoStop}%
\bibitem [{\citenamefont {{Abdikamalov}}\ \emph {et~al.}(2022)\citenamefont
  {{Abdikamalov}}, \citenamefont {{Pagliaroli}},\ and\ \citenamefont
  {{Radice}}}]{AbPaRa22}%
  \BibitemOpen
  \bibfield  {author} {\bibinfo {author} {\bibfnamefont {E.}~\bibnamefont
  {{Abdikamalov}}}, \bibinfo {author} {\bibfnamefont {G.}~\bibnamefont
  {{Pagliaroli}}}, \ and\ \bibinfo {author} {\bibfnamefont {D.}~\bibnamefont
  {{Radice}}},\ }in\ \href {\doibase 10.1007/978-981-15-4702-7_21-1} {\emph
  {\bibinfo {booktitle} {Handbook of Gravitational Wave Astronomy. Edited by C.
  Bambi}}}\ (\bibinfo {year} {2022})\ p.~\bibinfo {pages} {21}\BibitemShut
  {NoStop}%
\bibitem [{\citenamefont {Aasi}\ \emph {et~al.}(2015)\citenamefont {Aasi} \emph
  {et~al.}}]{LIGOScientific:2014pky}%
  \BibitemOpen
  \bibfield  {author} {\bibinfo {author} {\bibfnamefont {J.}~\bibnamefont
  {Aasi}} \emph {et~al.} (\bibinfo {collaboration} {LIGO Scientific}),\ }\href
  {\doibase 10.1088/0264-9381/32/7/074001} {\bibfield  {journal} {\bibinfo
  {journal} {Classical and Quantum Gravity.}\ }\textbf {\bibinfo {volume}
  {32}},\ \bibinfo {pages} {074001} (\bibinfo {year} {2015})},\ \Eprint
  {http://arxiv.org/abs/1411.4547} {arXiv:1411.4547 [gr-qc]} \BibitemShut
  {NoStop}%
\bibitem [{\citenamefont {Acernese}(2014)}]{Acernese_2014}%
  \BibitemOpen
  \bibfield  {author} {\bibinfo {author} {\bibfnamefont {F.}~\bibnamefont
  {Acernese}},\ }\href {\doibase 10.1088/0264-9381/32/2/024001} {\bibfield
  {journal} {\bibinfo  {journal} {Classical and Quantum Gravity}\ }\textbf
  {\bibinfo {volume} {32}},\ \bibinfo {pages} {024001} (\bibinfo {year}
  {2014})}\BibitemShut {NoStop}%
\bibitem [{\citenamefont {Aso}\ \emph {et~al.}(2013)\citenamefont {Aso},
  \citenamefont {Michimura}, \citenamefont {Somiya}, \citenamefont {Ando},
  \citenamefont {Miyakawa}, \citenamefont {Sekiguchi}, \citenamefont
  {Tatsumi},\ and\ \citenamefont {Yamamoto}}]{Aso_2013}%
  \BibitemOpen
  \bibfield  {author} {\bibinfo {author} {\bibfnamefont {Y.}~\bibnamefont
  {Aso}}, \bibinfo {author} {\bibfnamefont {Y.}~\bibnamefont {Michimura}},
  \bibinfo {author} {\bibfnamefont {K.}~\bibnamefont {Somiya}}, \bibinfo
  {author} {\bibfnamefont {M.}~\bibnamefont {Ando}}, \bibinfo {author}
  {\bibfnamefont {O.}~\bibnamefont {Miyakawa}}, \bibinfo {author}
  {\bibfnamefont {T.}~\bibnamefont {Sekiguchi}}, \bibinfo {author}
  {\bibfnamefont {D.}~\bibnamefont {Tatsumi}}, \ and\ \bibinfo {author}
  {\bibfnamefont {H.}~\bibnamefont {Yamamoto}},\ }\href {\doibase
  10.1103/physrevd.88.043007} {\bibfield  {journal} {\bibinfo  {journal}
  {Physical Review D}\ }\textbf {\bibinfo {volume} {88}} (\bibinfo {year}
  {2013}),\ 10.1103/physrevd.88.043007}\BibitemShut {NoStop}%
\bibitem [{\citenamefont {{Janka}}\ \emph {et~al.}(2016)\citenamefont
  {{Janka}}, \citenamefont {{Melson}},\ and\ \citenamefont
  {{Summa}}}]{JaMeSu16}%
  \BibitemOpen
  \bibfield  {author} {\bibinfo {author} {\bibfnamefont {H.-T.}\ \bibnamefont
  {{Janka}}}, \bibinfo {author} {\bibfnamefont {T.}~\bibnamefont {{Melson}}}, \
  and\ \bibinfo {author} {\bibfnamefont {A.}~\bibnamefont {{Summa}}},\ }\href
  {\doibase 10.1146/annurev-nucl-102115-044747} {\bibfield  {journal} {\bibinfo
   {journal} {Annual Review of Nuclear and Particle Science}\ }\textbf
  {\bibinfo {volume} {66}},\ \bibinfo {pages} {341} (\bibinfo {year} {2016})},\
  \Eprint {http://arxiv.org/abs/1602.05576} {arXiv:1602.05576 [astro-ph.SR]}
  \BibitemShut {NoStop}%
\bibitem [{\citenamefont {{M{\"u}ller}}(2016)}]{Mueller16}%
  \BibitemOpen
  \bibfield  {author} {\bibinfo {author} {\bibfnamefont {B.}~\bibnamefont
  {{M{\"u}ller}}},\ }\href {\doibase 10.1017/pasa.2016.40} {\bibfield
  {journal} {\bibinfo  {journal} {\pasa}\ }\textbf {\bibinfo {volume} {33}},\
  \bibinfo {eid} {e048} (\bibinfo {year} {2016})},\ \Eprint
  {http://arxiv.org/abs/1608.03274} {arXiv:1608.03274 [astro-ph.SR]}
  \BibitemShut {NoStop}%
\bibitem [{\citenamefont {{Mezzacappa}}\ \emph {et~al.}(2020)\citenamefont
  {{Mezzacappa}}, \citenamefont {{Endeve}}, \citenamefont {{Messer}},\ and\
  \citenamefont {{Bruenn}}}]{MeEnMe20}%
  \BibitemOpen
  \bibfield  {author} {\bibinfo {author} {\bibfnamefont {A.}~\bibnamefont
  {{Mezzacappa}}}, \bibinfo {author} {\bibfnamefont {E.}~\bibnamefont
  {{Endeve}}}, \bibinfo {author} {\bibfnamefont {O.~E.~B.}\ \bibnamefont
  {{Messer}}}, \ and\ \bibinfo {author} {\bibfnamefont {S.~W.}\ \bibnamefont
  {{Bruenn}}},\ }\href {\doibase 10.1007/s41115-020-00010-8} {\bibfield
  {journal} {\bibinfo  {journal} {Living Reviews in Computational
  Astrophysics}\ }\textbf {\bibinfo {volume} {6}},\ \bibinfo {eid} {4}
  (\bibinfo {year} {2020})},\ \Eprint {http://arxiv.org/abs/2010.09013}
  {arXiv:2010.09013 [astro-ph.HE]} \BibitemShut {NoStop}%
\bibitem [{\citenamefont {{M{\"u}ller}}(2020)}]{Mueller20}%
  \BibitemOpen
  \bibfield  {author} {\bibinfo {author} {\bibfnamefont {B.}~\bibnamefont
  {{M{\"u}ller}}},\ }\href {\doibase 10.1007/s41115-020-0008-5} {\bibfield
  {journal} {\bibinfo  {journal} {Living Reviews in Computational
  Astrophysics}\ }\textbf {\bibinfo {volume} {6}},\ \bibinfo {eid} {3}
  (\bibinfo {year} {2020})},\ \Eprint {http://arxiv.org/abs/2006.05083}
  {arXiv:2006.05083 [astro-ph.SR]} \BibitemShut {NoStop}%
\bibitem [{\citenamefont {{Burrows}}\ and\ \citenamefont
  {{Vartanyan}}(2021)}]{BuVa21}%
  \BibitemOpen
  \bibfield  {author} {\bibinfo {author} {\bibfnamefont {A.}~\bibnamefont
  {{Burrows}}}\ and\ \bibinfo {author} {\bibfnamefont {D.}~\bibnamefont
  {{Vartanyan}}},\ }\href {\doibase 10.1038/s41586-020-03059-w} {\bibfield
  {journal} {\bibinfo  {journal} {\nat}\ }\textbf {\bibinfo {volume} {589}},\
  \bibinfo {pages} {29} (\bibinfo {year} {2021})},\ \Eprint
  {http://arxiv.org/abs/2009.14157} {arXiv:2009.14157 [astro-ph.SR]}
  \BibitemShut {NoStop}%
\bibitem [{\citenamefont {{Mezzacappa}}(2023)}]{Mezzacappa23}%
  \BibitemOpen
  \bibfield  {author} {\bibinfo {author} {\bibfnamefont {A.}~\bibnamefont
  {{Mezzacappa}}},\ }\href {\doibase 10.1017/S1743921322001831} {\bibfield
  {journal} {\bibinfo  {journal} {IAU Symposium}\ }\textbf {\bibinfo {volume}
  {362}},\ \bibinfo {pages} {215} (\bibinfo {year} {2023})},\ \Eprint
  {http://arxiv.org/abs/2205.13438} {arXiv:2205.13438 [astro-ph.SR]}
  \BibitemShut {NoStop}%
\bibitem [{\citenamefont {Andresen}\ \emph {et~al.}(2017)\citenamefont
  {Andresen}, \citenamefont {Müller}, \citenamefont {Müller},\ and\
  \citenamefont {Janka}}]{Andresen_2017}%
  \BibitemOpen
  \bibfield  {author} {\bibinfo {author} {\bibfnamefont {H.}~\bibnamefont
  {Andresen}}, \bibinfo {author} {\bibfnamefont {B.}~\bibnamefont {Müller}},
  \bibinfo {author} {\bibfnamefont {E.}~\bibnamefont {Müller}}, \ and\
  \bibinfo {author} {\bibfnamefont {H.-T.}\ \bibnamefont {Janka}},\ }\href
  {\doibase 10.1093/mnras/stx618} {\bibfield  {journal} {\bibinfo  {journal}
  {Monthly Notices of the Royal Astronomical Society}\ }\textbf {\bibinfo
  {volume} {468}},\ \bibinfo {pages} {2032} (\bibinfo {year}
  {2017})}\BibitemShut {NoStop}%
\bibitem [{\citenamefont {Mezzacappa}\ \emph {et~al.}(2020)\citenamefont
  {Mezzacappa}, \citenamefont {Marronetti}, \citenamefont {Landfield},
  \citenamefont {Lentz}, \citenamefont {Yakunin}, \citenamefont {Bruenn},
  \citenamefont {Hix}, \citenamefont {Messer}, \citenamefont {Endeve},
  \citenamefont {Blondin},\ and\ \citenamefont {Harris}}]{Mezzacappa_2020}%
  \BibitemOpen
  \bibfield  {author} {\bibinfo {author} {\bibfnamefont {A.}~\bibnamefont
  {Mezzacappa}}, \bibinfo {author} {\bibfnamefont {P.}~\bibnamefont
  {Marronetti}}, \bibinfo {author} {\bibfnamefont {R.~E.}\ \bibnamefont
  {Landfield}}, \bibinfo {author} {\bibfnamefont {E.~J.}\ \bibnamefont
  {Lentz}}, \bibinfo {author} {\bibfnamefont {K.~N.}\ \bibnamefont {Yakunin}},
  \bibinfo {author} {\bibfnamefont {S.~W.}\ \bibnamefont {Bruenn}}, \bibinfo
  {author} {\bibfnamefont {W.~R.}\ \bibnamefont {Hix}}, \bibinfo {author}
  {\bibfnamefont {O.~B.}\ \bibnamefont {Messer}}, \bibinfo {author}
  {\bibfnamefont {E.}~\bibnamefont {Endeve}}, \bibinfo {author} {\bibfnamefont
  {J.~M.}\ \bibnamefont {Blondin}}, \ and\ \bibinfo {author} {\bibfnamefont
  {J.~A.}\ \bibnamefont {Harris}},\ }\href {\doibase
  10.1103/physrevd.102.023027} {\bibfield  {journal} {\bibinfo  {journal}
  {Physical Review D}\ }\textbf {\bibinfo {volume} {102}} (\bibinfo {year}
  {2020}),\ 10.1103/physrevd.102.023027}\BibitemShut {NoStop}%
\bibitem [{\citenamefont {{Mezzacappa}}\ \emph {et~al.}(2023)\citenamefont
  {{Mezzacappa}}, \citenamefont {{Marronetti}}, \citenamefont {{Landfield}},
  \citenamefont {{Lentz}}, \citenamefont {{Murphy}}, \citenamefont {{Raphael
  Hix}}, \citenamefont {{Harris}}, \citenamefont {{Bruenn}}, \citenamefont
  {{Blondin}}, \citenamefont {{Bronson Messer}}, \citenamefont {{Casanova}},\
  and\ \citenamefont {{Kronzer}}}]{MeMaLa23}%
  \BibitemOpen
  \bibfield  {author} {\bibinfo {author} {\bibfnamefont {A.}~\bibnamefont
  {{Mezzacappa}}}, \bibinfo {author} {\bibfnamefont {P.}~\bibnamefont
  {{Marronetti}}}, \bibinfo {author} {\bibfnamefont {R.~E.}\ \bibnamefont
  {{Landfield}}}, \bibinfo {author} {\bibfnamefont {E.~J.}\ \bibnamefont
  {{Lentz}}}, \bibinfo {author} {\bibfnamefont {R.~D.}\ \bibnamefont
  {{Murphy}}}, \bibinfo {author} {\bibfnamefont {W.}~\bibnamefont {{Raphael
  Hix}}}, \bibinfo {author} {\bibfnamefont {J.~A.}\ \bibnamefont {{Harris}}},
  \bibinfo {author} {\bibfnamefont {S.~W.}\ \bibnamefont {{Bruenn}}}, \bibinfo
  {author} {\bibfnamefont {J.~M.}\ \bibnamefont {{Blondin}}}, \bibinfo {author}
  {\bibfnamefont {O.~E.}\ \bibnamefont {{Bronson Messer}}}, \bibinfo {author}
  {\bibfnamefont {J.}~\bibnamefont {{Casanova}}}, \ and\ \bibinfo {author}
  {\bibfnamefont {L.~L.}\ \bibnamefont {{Kronzer}}},\ }\href {\doibase
  10.1103/PhysRevD.107.043008} {\bibfield  {journal} {\bibinfo  {journal}
  {\prd}\ }\textbf {\bibinfo {volume} {107}},\ \bibinfo {eid} {043008}
  (\bibinfo {year} {2023})},\ \Eprint {http://arxiv.org/abs/2208.10643}
  {arXiv:2208.10643 [astro-ph.SR]} \BibitemShut {NoStop}%
\bibitem [{\citenamefont {Andresen}\ \emph {et~al.}(2019)\citenamefont
  {Andresen}, \citenamefont {M\"uller}, \citenamefont {Janka}, \citenamefont
  {Summa}, \citenamefont {Gill},\ and\ \citenamefont
  {Zanolin}}]{Andresen:2018aom}%
  \BibitemOpen
  \bibfield  {author} {\bibinfo {author} {\bibfnamefont {H.}~\bibnamefont
  {Andresen}}, \bibinfo {author} {\bibfnamefont {E.}~\bibnamefont {M\"uller}},
  \bibinfo {author} {\bibfnamefont {H.~T.}\ \bibnamefont {Janka}}, \bibinfo
  {author} {\bibfnamefont {A.}~\bibnamefont {Summa}}, \bibinfo {author}
  {\bibfnamefont {K.}~\bibnamefont {Gill}}, \ and\ \bibinfo {author}
  {\bibfnamefont {M.}~\bibnamefont {Zanolin}},\ }\href {\doibase
  10.1093/mnras/stz990} {\bibfield  {journal} {\bibinfo  {journal} {Mon. Not.
  Roy. Astron. Soc.}\ }\textbf {\bibinfo {volume} {486}},\ \bibinfo {pages}
  {2238} (\bibinfo {year} {2019})},\ \Eprint {http://arxiv.org/abs/1810.07638}
  {arXiv:1810.07638 [astro-ph.HE]} \BibitemShut {NoStop}%
\bibitem [{\citenamefont {Cerd{\'{a} }-Dur{\'{a}}n}\ \emph
  {et~al.}(2013)\citenamefont {Cerd{\'{a} }-Dur{\'{a}}n}, \citenamefont
  {DeBrye}, \citenamefont {Aloy}, \citenamefont {Font},\ and\ \citenamefont
  {Obergaulinger}}]{Cerd_Dur_n_2013}%
  \BibitemOpen
  \bibfield  {author} {\bibinfo {author} {\bibfnamefont {P.}~\bibnamefont
  {Cerd{\'{a} }-Dur{\'{a}}n}}, \bibinfo {author} {\bibfnamefont
  {N.}~\bibnamefont {DeBrye}}, \bibinfo {author} {\bibfnamefont {M.~A.}\
  \bibnamefont {Aloy}}, \bibinfo {author} {\bibfnamefont {J.~A.}\ \bibnamefont
  {Font}}, \ and\ \bibinfo {author} {\bibfnamefont {M.}~\bibnamefont
  {Obergaulinger}},\ }\href {\doibase 10.1088/2041-8205/779/2/l18} {\bibfield
  {journal} {\bibinfo  {journal} {The Astrophysical Journal}\ }\textbf
  {\bibinfo {volume} {779}},\ \bibinfo {pages} {L18} (\bibinfo {year}
  {2013})}\BibitemShut {NoStop}%
\bibitem [{\citenamefont {{Hayama}}\ \emph {et~al.}(2015)\citenamefont
  {{Hayama}}, \citenamefont {{Kuroda}}, \citenamefont {{Kotake}},\ and\
  \citenamefont {{Takiwaki}}}]{HaKuKo15}%
  \BibitemOpen
  \bibfield  {author} {\bibinfo {author} {\bibfnamefont {K.}~\bibnamefont
  {{Hayama}}}, \bibinfo {author} {\bibfnamefont {T.}~\bibnamefont {{Kuroda}}},
  \bibinfo {author} {\bibfnamefont {K.}~\bibnamefont {{Kotake}}}, \ and\
  \bibinfo {author} {\bibfnamefont {T.}~\bibnamefont {{Takiwaki}}},\ }\href
  {\doibase 10.1103/PhysRevD.92.122001} {\bibfield  {journal} {\bibinfo
  {journal} {\prd}\ }\textbf {\bibinfo {volume} {92}},\ \bibinfo {eid} {122001}
  (\bibinfo {year} {2015})},\ \Eprint {http://arxiv.org/abs/1501.00966}
  {arXiv:1501.00966 [astro-ph.HE]} \BibitemShut {NoStop}%
\bibitem [{\citenamefont {{Hayama}}\ \emph {et~al.}(2018)\citenamefont
  {{Hayama}}, \citenamefont {{Kuroda}}, \citenamefont {{Kotake}},\ and\
  \citenamefont {{Takiwaki}}}]{HaKuKo18}%
  \BibitemOpen
  \bibfield  {author} {\bibinfo {author} {\bibfnamefont {K.}~\bibnamefont
  {{Hayama}}}, \bibinfo {author} {\bibfnamefont {T.}~\bibnamefont {{Kuroda}}},
  \bibinfo {author} {\bibfnamefont {K.}~\bibnamefont {{Kotake}}}, \ and\
  \bibinfo {author} {\bibfnamefont {T.}~\bibnamefont {{Takiwaki}}},\ }\href
  {\doibase 10.1093/mnrasl/sly055} {\bibfield  {journal} {\bibinfo  {journal}
  {\mnras}\ }\textbf {\bibinfo {volume} {477}},\ \bibinfo {pages} {L96}
  (\bibinfo {year} {2018})},\ \Eprint {http://arxiv.org/abs/1802.03842}
  {arXiv:1802.03842 [astro-ph.HE]} \BibitemShut {NoStop}%
\bibitem [{\citenamefont {{Hayama}}\ \emph {et~al.}(2016)\citenamefont
  {{Hayama}}, \citenamefont {{Kuroda}}, \citenamefont {{Nakamura}},\ and\
  \citenamefont {{Yamada}}}]{HaKuNa16}%
  \BibitemOpen
  \bibfield  {author} {\bibinfo {author} {\bibfnamefont {K.}~\bibnamefont
  {{Hayama}}}, \bibinfo {author} {\bibfnamefont {T.}~\bibnamefont {{Kuroda}}},
  \bibinfo {author} {\bibfnamefont {K.}~\bibnamefont {{Nakamura}}}, \ and\
  \bibinfo {author} {\bibfnamefont {S.}~\bibnamefont {{Yamada}}},\ }\href
  {\doibase 10.1103/PhysRevLett.116.151102} {\bibfield  {journal} {\bibinfo
  {journal} {\prl}\ }\textbf {\bibinfo {volume} {116}},\ \bibinfo {eid}
  {151102} (\bibinfo {year} {2016})},\ \Eprint
  {http://arxiv.org/abs/1606.01520} {arXiv:1606.01520 [astro-ph.HE]}
  \BibitemShut {NoStop}%
\bibitem [{\citenamefont {{Kawahara}}\ \emph {et~al.}(2018)\citenamefont
  {{Kawahara}}, \citenamefont {{Kuroda}}, \citenamefont {{Takiwaki}},
  \citenamefont {{Hayama}},\ and\ \citenamefont {{Kotake}}}]{KaKuTa18}%
  \BibitemOpen
  \bibfield  {author} {\bibinfo {author} {\bibfnamefont {H.}~\bibnamefont
  {{Kawahara}}}, \bibinfo {author} {\bibfnamefont {T.}~\bibnamefont
  {{Kuroda}}}, \bibinfo {author} {\bibfnamefont {T.}~\bibnamefont
  {{Takiwaki}}}, \bibinfo {author} {\bibfnamefont {K.}~\bibnamefont
  {{Hayama}}}, \ and\ \bibinfo {author} {\bibfnamefont {K.}~\bibnamefont
  {{Kotake}}},\ }\href {\doibase 10.3847/1538-4357/aae57b} {\bibfield
  {journal} {\bibinfo  {journal} {\apj}\ }\textbf {\bibinfo {volume} {867}},\
  \bibinfo {eid} {126} (\bibinfo {year} {2018})},\ \Eprint
  {http://arxiv.org/abs/1810.00334} {arXiv:1810.00334 [astro-ph.HE]}
  \BibitemShut {NoStop}%
\bibitem [{\citenamefont {{Kotake}}\ \emph {et~al.}(2011)\citenamefont
  {{Kotake}}, \citenamefont {{Iwakami-Nakano}},\ and\ \citenamefont
  {{Ohnishi}}}]{KoIwOh11}%
  \BibitemOpen
  \bibfield  {author} {\bibinfo {author} {\bibfnamefont {K.}~\bibnamefont
  {{Kotake}}}, \bibinfo {author} {\bibfnamefont {W.}~\bibnamefont
  {{Iwakami-Nakano}}}, \ and\ \bibinfo {author} {\bibfnamefont
  {N.}~\bibnamefont {{Ohnishi}}},\ }\href {\doibase
  10.1088/0004-637X/736/2/124} {\bibfield  {journal} {\bibinfo  {journal}
  {\apj}\ }\textbf {\bibinfo {volume} {736}},\ \bibinfo {eid} {124} (\bibinfo
  {year} {2011})},\ \Eprint {http://arxiv.org/abs/1106.0544} {arXiv:1106.0544
  [astro-ph.HE]} \BibitemShut {NoStop}%
\bibitem [{\citenamefont {{Kotake}}\ \emph {et~al.}(2009)\citenamefont
  {{Kotake}}, \citenamefont {{Iwakami}}, \citenamefont {{Ohnishi}},\ and\
  \citenamefont {{Yamada}}}]{KoIwOh09}%
  \BibitemOpen
  \bibfield  {author} {\bibinfo {author} {\bibfnamefont {K.}~\bibnamefont
  {{Kotake}}}, \bibinfo {author} {\bibfnamefont {W.}~\bibnamefont {{Iwakami}}},
  \bibinfo {author} {\bibfnamefont {N.}~\bibnamefont {{Ohnishi}}}, \ and\
  \bibinfo {author} {\bibfnamefont {S.}~\bibnamefont {{Yamada}}},\ }\href
  {\doibase 10.1088/0004-637X/697/2/L133} {\bibfield  {journal} {\bibinfo
  {journal} {\apjl}\ }\textbf {\bibinfo {volume} {697}},\ \bibinfo {pages}
  {L133} (\bibinfo {year} {2009})},\ \Eprint {http://arxiv.org/abs/0904.4300}
  {arXiv:0904.4300 [astro-ph.HE]} \BibitemShut {NoStop}%
\bibitem [{\citenamefont {{Kuroda}}\ \emph {et~al.}(2022)\citenamefont
  {{Kuroda}}, \citenamefont {{Fischer}}, \citenamefont {{Takiwaki}},\ and\
  \citenamefont {{Kotake}}}]{KuFiTa22}%
  \BibitemOpen
  \bibfield  {author} {\bibinfo {author} {\bibfnamefont {T.}~\bibnamefont
  {{Kuroda}}}, \bibinfo {author} {\bibfnamefont {T.}~\bibnamefont {{Fischer}}},
  \bibinfo {author} {\bibfnamefont {T.}~\bibnamefont {{Takiwaki}}}, \ and\
  \bibinfo {author} {\bibfnamefont {K.}~\bibnamefont {{Kotake}}},\ }\href
  {\doibase 10.3847/1538-4357/ac31a8} {\bibfield  {journal} {\bibinfo
  {journal} {\apj}\ }\textbf {\bibinfo {volume} {924}},\ \bibinfo {eid} {38}
  (\bibinfo {year} {2022})},\ \Eprint {http://arxiv.org/abs/2109.01508}
  {arXiv:2109.01508 [astro-ph.HE]} \BibitemShut {NoStop}%
\bibitem [{\citenamefont {Kuroda}\ \emph {et~al.}(2017)\citenamefont {Kuroda},
  \citenamefont {Kotake}, \citenamefont {Hayama},\ and\ \citenamefont
  {Takiwaki}}]{Kuroda_2017}%
  \BibitemOpen
  \bibfield  {author} {\bibinfo {author} {\bibfnamefont {T.}~\bibnamefont
  {Kuroda}}, \bibinfo {author} {\bibfnamefont {K.}~\bibnamefont {Kotake}},
  \bibinfo {author} {\bibfnamefont {K.}~\bibnamefont {Hayama}}, \ and\ \bibinfo
  {author} {\bibfnamefont {T.}~\bibnamefont {Takiwaki}},\ }\href {\doibase
  10.3847/1538-4357/aa988d} {\bibfield  {journal} {\bibinfo  {journal} {The
  Astrophysical Journal}\ }\textbf {\bibinfo {volume} {851}},\ \bibinfo {pages}
  {62} (\bibinfo {year} {2017})}\BibitemShut {NoStop}%
\bibitem [{\citenamefont {{Kuroda}}\ \emph {et~al.}(2016)\citenamefont
  {{Kuroda}}, \citenamefont {{Kotake}},\ and\ \citenamefont
  {{Takiwaki}}}]{KuKoTa16}%
  \BibitemOpen
  \bibfield  {author} {\bibinfo {author} {\bibfnamefont {T.}~\bibnamefont
  {{Kuroda}}}, \bibinfo {author} {\bibfnamefont {K.}~\bibnamefont {{Kotake}}},
  \ and\ \bibinfo {author} {\bibfnamefont {T.}~\bibnamefont {{Takiwaki}}},\
  }\href {\doibase 10.3847/2041-8205/829/1/L14} {\bibfield  {journal} {\bibinfo
   {journal} {\apjl}\ }\textbf {\bibinfo {volume} {829}},\ \bibinfo {eid} {L14}
  (\bibinfo {year} {2016})},\ \Eprint {http://arxiv.org/abs/1605.09215}
  {arXiv:1605.09215 [astro-ph.HE]} \BibitemShut {NoStop}%
\bibitem [{\citenamefont {Kuroda}\ \emph {et~al.}(2018)\citenamefont {Kuroda},
  \citenamefont {Kotake}, \citenamefont {Takiwaki},\ and\ \citenamefont
  {Thielemann}}]{Kuroda_2018}%
  \BibitemOpen
  \bibfield  {author} {\bibinfo {author} {\bibfnamefont {T.}~\bibnamefont
  {Kuroda}}, \bibinfo {author} {\bibfnamefont {K.}~\bibnamefont {Kotake}},
  \bibinfo {author} {\bibfnamefont {T.}~\bibnamefont {Takiwaki}}, \ and\
  \bibinfo {author} {\bibfnamefont {F.-K.}\ \bibnamefont {Thielemann}},\ }\href
  {\doibase 10.1093/mnrasl/sly059} {\bibfield  {journal} {\bibinfo  {journal}
  {Monthly Notices of the Royal Astronomical Society: Letters}\ }\textbf
  {\bibinfo {volume} {477}},\ \bibinfo {pages} {L80} (\bibinfo {year}
  {2018})}\BibitemShut {NoStop}%
\bibitem [{\citenamefont {Morozova}\ \emph {et~al.}(2018)\citenamefont
  {Morozova}, \citenamefont {Radice}, \citenamefont {Burrows},\ and\
  \citenamefont {Vartanyan}}]{Morozova_2018}%
  \BibitemOpen
  \bibfield  {author} {\bibinfo {author} {\bibfnamefont {V.}~\bibnamefont
  {Morozova}}, \bibinfo {author} {\bibfnamefont {D.}~\bibnamefont {Radice}},
  \bibinfo {author} {\bibfnamefont {A.}~\bibnamefont {Burrows}}, \ and\
  \bibinfo {author} {\bibfnamefont {D.}~\bibnamefont {Vartanyan}},\ }\href
  {\doibase 10.3847/1538-4357/aac5f1} {\bibfield  {journal} {\bibinfo
  {journal} {The Astrophysical Journal}\ }\textbf {\bibinfo {volume} {861}},\
  \bibinfo {pages} {10} (\bibinfo {year} {2018})}\BibitemShut {NoStop}%
\bibitem [{\citenamefont {Müller}\ \emph {et~al.}(2013)\citenamefont
  {Müller}, \citenamefont {Janka},\ and\ \citenamefont {Marek}}]{M_ller_2013}%
  \BibitemOpen
  \bibfield  {author} {\bibinfo {author} {\bibfnamefont {B.}~\bibnamefont
  {Müller}}, \bibinfo {author} {\bibfnamefont {H.-T.}\ \bibnamefont {Janka}},
  \ and\ \bibinfo {author} {\bibfnamefont {A.}~\bibnamefont {Marek}},\ }\href
  {\doibase 10.1088/0004-637x/766/1/43} {\bibfield  {journal} {\bibinfo
  {journal} {The Astrophysical Journal}\ }\textbf {\bibinfo {volume} {766}},\
  \bibinfo {pages} {43} (\bibinfo {year} {2013})}\BibitemShut {NoStop}%
\bibitem [{\citenamefont {Murphy}\ \emph {et~al.}(2009)\citenamefont {Murphy},
  \citenamefont {Ott},\ and\ \citenamefont {Burrows}}]{Murphy_2009}%
  \BibitemOpen
  \bibfield  {author} {\bibinfo {author} {\bibfnamefont {J.~W.}\ \bibnamefont
  {Murphy}}, \bibinfo {author} {\bibfnamefont {C.~D.}\ \bibnamefont {Ott}}, \
  and\ \bibinfo {author} {\bibfnamefont {A.}~\bibnamefont {Burrows}},\ }\href
  {\doibase 10.1088/0004-637x/707/2/1173} {\bibfield  {journal} {\bibinfo
  {journal} {The Astrophysical Journal}\ }\textbf {\bibinfo {volume} {707}},\
  \bibinfo {pages} {1173} (\bibinfo {year} {2009})}\BibitemShut {NoStop}%
\bibitem [{\citenamefont {Nakamura}\ \emph {et~al.}(2022)\citenamefont
  {Nakamura}, \citenamefont {Takiwaki},\ and\ \citenamefont
  {Kotake}}]{Nakamura_2022}%
  \BibitemOpen
  \bibfield  {author} {\bibinfo {author} {\bibfnamefont {K.}~\bibnamefont
  {Nakamura}}, \bibinfo {author} {\bibfnamefont {T.}~\bibnamefont {Takiwaki}},
  \ and\ \bibinfo {author} {\bibfnamefont {K.}~\bibnamefont {Kotake}},\ }\href
  {\doibase 10.1093/mnras/stac1586} {\bibfield  {journal} {\bibinfo  {journal}
  {Monthly Notices of the Royal Astronomical Society}\ }\textbf {\bibinfo
  {volume} {514}},\ \bibinfo {pages} {3941} (\bibinfo {year}
  {2022})}\BibitemShut {NoStop}%
\bibitem [{\citenamefont {O'Connor}\ and\ \citenamefont
  {Couch}(2018)}]{O_Connor_2018}%
  \BibitemOpen
  \bibfield  {author} {\bibinfo {author} {\bibfnamefont {E.~P.}\ \bibnamefont
  {O'Connor}}\ and\ \bibinfo {author} {\bibfnamefont {S.~M.}\ \bibnamefont
  {Couch}},\ }\href {\doibase 10.3847/1538-4357/aadcf7} {\bibfield  {journal}
  {\bibinfo  {journal} {The Astrophysical Journal}\ }\textbf {\bibinfo {volume}
  {865}},\ \bibinfo {pages} {81} (\bibinfo {year} {2018})}\BibitemShut
  {NoStop}%
\bibitem [{\citenamefont {Ott}\ \emph {et~al.}(2013)\citenamefont {Ott},
  \citenamefont {Abdikamalov}, \citenamefont {Mösta}, \citenamefont {Haas},
  \citenamefont {Drasco}, \citenamefont {O{\textquotesingle}Connor},
  \citenamefont {Reisswig}, \citenamefont {Meakin},\ and\ \citenamefont
  {Schnetter}}]{Ott_2013}%
  \BibitemOpen
  \bibfield  {author} {\bibinfo {author} {\bibfnamefont {C.~D.}\ \bibnamefont
  {Ott}}, \bibinfo {author} {\bibfnamefont {E.}~\bibnamefont {Abdikamalov}},
  \bibinfo {author} {\bibfnamefont {P.}~\bibnamefont {Mösta}}, \bibinfo
  {author} {\bibfnamefont {R.}~\bibnamefont {Haas}}, \bibinfo {author}
  {\bibfnamefont {S.}~\bibnamefont {Drasco}}, \bibinfo {author} {\bibfnamefont
  {E.~P.}\ \bibnamefont {O{\textquotesingle}Connor}}, \bibinfo {author}
  {\bibfnamefont {C.}~\bibnamefont {Reisswig}}, \bibinfo {author}
  {\bibfnamefont {C.~A.}\ \bibnamefont {Meakin}}, \ and\ \bibinfo {author}
  {\bibfnamefont {E.}~\bibnamefont {Schnetter}},\ }\href {\doibase
  10.1088/0004-637x/768/2/115} {\bibfield  {journal} {\bibinfo  {journal} {The
  Astrophysical Journal}\ }\textbf {\bibinfo {volume} {768}},\ \bibinfo {pages}
  {115} (\bibinfo {year} {2013})}\BibitemShut {NoStop}%
\bibitem [{\citenamefont {Pajkos}\ \emph {et~al.}(2019)\citenamefont {Pajkos},
  \citenamefont {Couch}, \citenamefont {Pan},\ and\ \citenamefont
  {O'Connor}}]{Pajkos_2019}%
  \BibitemOpen
  \bibfield  {author} {\bibinfo {author} {\bibfnamefont {M.~A.}\ \bibnamefont
  {Pajkos}}, \bibinfo {author} {\bibfnamefont {S.~M.}\ \bibnamefont {Couch}},
  \bibinfo {author} {\bibfnamefont {K.-C.}\ \bibnamefont {Pan}}, \ and\
  \bibinfo {author} {\bibfnamefont {E.~P.}\ \bibnamefont {O'Connor}},\ }\href
  {\doibase 10.3847/1538-4357/ab1de2} {\bibfield  {journal} {\bibinfo
  {journal} {The Astrophysical Journal}\ }\textbf {\bibinfo {volume} {878}},\
  \bibinfo {pages} {13} (\bibinfo {year} {2019})}\BibitemShut {NoStop}%
\bibitem [{\citenamefont {{Pajkos}}\ \emph {et~al.}(2021)\citenamefont
  {{Pajkos}}, \citenamefont {{Warren}}, \citenamefont {{Couch}}, \citenamefont
  {{O'Connor}},\ and\ \citenamefont {{Pan}}}]{PaWaCo21}%
  \BibitemOpen
  \bibfield  {author} {\bibinfo {author} {\bibfnamefont {M.~A.}\ \bibnamefont
  {{Pajkos}}}, \bibinfo {author} {\bibfnamefont {M.~L.}\ \bibnamefont
  {{Warren}}}, \bibinfo {author} {\bibfnamefont {S.~M.}\ \bibnamefont
  {{Couch}}}, \bibinfo {author} {\bibfnamefont {E.~P.}\ \bibnamefont
  {{O'Connor}}}, \ and\ \bibinfo {author} {\bibfnamefont {K.-C.}\ \bibnamefont
  {{Pan}}},\ }\href {\doibase 10.3847/1538-4357/abfb65} {\bibfield  {journal}
  {\bibinfo  {journal} {\apj}\ }\textbf {\bibinfo {volume} {914}},\ \bibinfo
  {eid} {80} (\bibinfo {year} {2021})},\ \Eprint
  {http://arxiv.org/abs/2011.09000} {arXiv:2011.09000 [astro-ph.HE]}
  \BibitemShut {NoStop}%
\bibitem [{\citenamefont {Pan}\ \emph {et~al.}(2018)\citenamefont {Pan},
  \citenamefont {Liebendörfer}, \citenamefont {Couch},\ and\ \citenamefont
  {Thielemann}}]{Pan_2018}%
  \BibitemOpen
  \bibfield  {author} {\bibinfo {author} {\bibfnamefont {K.-C.}\ \bibnamefont
  {Pan}}, \bibinfo {author} {\bibfnamefont {M.}~\bibnamefont {Liebendörfer}},
  \bibinfo {author} {\bibfnamefont {S.~M.}\ \bibnamefont {Couch}}, \ and\
  \bibinfo {author} {\bibfnamefont {F.-K.}\ \bibnamefont {Thielemann}},\ }\href
  {\doibase 10.3847/1538-4357/aab71d} {\bibfield  {journal} {\bibinfo
  {journal} {The Astrophysical Journal}\ }\textbf {\bibinfo {volume} {857}},\
  \bibinfo {pages} {13} (\bibinfo {year} {2018})}\BibitemShut {NoStop}%
\bibitem [{\citenamefont {Pan}\ \emph {et~al.}(2021)\citenamefont {Pan},
  \citenamefont {Liebendörfer}, \citenamefont {Couch},\ and\ \citenamefont
  {Thielemann}}]{Pan_2021}%
  \BibitemOpen
  \bibfield  {author} {\bibinfo {author} {\bibfnamefont {K.-C.}\ \bibnamefont
  {Pan}}, \bibinfo {author} {\bibfnamefont {M.}~\bibnamefont {Liebendörfer}},
  \bibinfo {author} {\bibfnamefont {S.~M.}\ \bibnamefont {Couch}}, \ and\
  \bibinfo {author} {\bibfnamefont {F.-K.}\ \bibnamefont {Thielemann}},\ }\href
  {\doibase 10.3847/1538-4357/abfb05} {\bibfield  {journal} {\bibinfo
  {journal} {The Astrophysical Journal}\ }\textbf {\bibinfo {volume} {914}},\
  \bibinfo {pages} {140} (\bibinfo {year} {2021})}\BibitemShut {NoStop}%
\bibitem [{\citenamefont {Powell}\ and\ \citenamefont
  {Müller}(2019)}]{Powell_2019}%
  \BibitemOpen
  \bibfield  {author} {\bibinfo {author} {\bibfnamefont {J.}~\bibnamefont
  {Powell}}\ and\ \bibinfo {author} {\bibfnamefont {B.}~\bibnamefont
  {Müller}},\ }\href {\doibase 10.1093/mnras/stz1304} {\bibfield  {journal}
  {\bibinfo  {journal} {Monthly Notices of the Royal Astronomical Society}\
  }\textbf {\bibinfo {volume} {487}},\ \bibinfo {pages} {1178} (\bibinfo {year}
  {2019})}\BibitemShut {NoStop}%
\bibitem [{\citenamefont {Powell}\ and\ \citenamefont
  {Müller}(2020)}]{Powell_2020}%
  \BibitemOpen
  \bibfield  {author} {\bibinfo {author} {\bibfnamefont {J.}~\bibnamefont
  {Powell}}\ and\ \bibinfo {author} {\bibfnamefont {B.}~\bibnamefont
  {Müller}},\ }\href {\doibase 10.1093/mnras/staa1048} {\bibfield  {journal}
  {\bibinfo  {journal} {Monthly Notices of the Royal Astronomical Society}\
  }\textbf {\bibinfo {volume} {494}},\ \bibinfo {pages} {4665} (\bibinfo {year}
  {2020})}\BibitemShut {NoStop}%
\bibitem [{\citenamefont {Powell}\ and\ \citenamefont
  {Müller}(2022)}]{Powell_2022}%
  \BibitemOpen
  \bibfield  {author} {\bibinfo {author} {\bibfnamefont {J.}~\bibnamefont
  {Powell}}\ and\ \bibinfo {author} {\bibfnamefont {B.}~\bibnamefont
  {Müller}},\ }\href {\doibase 10.1103/physrevd.105.063018} {\bibfield
  {journal} {\bibinfo  {journal} {Physical Review D}\ }\textbf {\bibinfo
  {volume} {105}} (\bibinfo {year} {2022}),\
  10.1103/physrevd.105.063018}\BibitemShut {NoStop}%
\bibitem [{\citenamefont {Radice}\ \emph {et~al.}(2019)\citenamefont {Radice},
  \citenamefont {Morozova}, \citenamefont {Burrows}, \citenamefont
  {Vartanyan},\ and\ \citenamefont {Nagakura}}]{Radice_2019}%
  \BibitemOpen
  \bibfield  {author} {\bibinfo {author} {\bibfnamefont {D.}~\bibnamefont
  {Radice}}, \bibinfo {author} {\bibfnamefont {V.}~\bibnamefont {Morozova}},
  \bibinfo {author} {\bibfnamefont {A.}~\bibnamefont {Burrows}}, \bibinfo
  {author} {\bibfnamefont {D.}~\bibnamefont {Vartanyan}}, \ and\ \bibinfo
  {author} {\bibfnamefont {H.}~\bibnamefont {Nagakura}},\ }\href {\doibase
  10.3847/2041-8213/ab191a} {\bibfield  {journal} {\bibinfo  {journal} {The
  Astrophysical Journal}\ }\textbf {\bibinfo {volume} {876}},\ \bibinfo {pages}
  {L9} (\bibinfo {year} {2019})}\BibitemShut {NoStop}%
\bibitem [{\citenamefont {{Richardson}}\ \emph {et~al.}(2022)\citenamefont
  {{Richardson}}, \citenamefont {{Zanolin}}, \citenamefont {{Andresen}},
  \citenamefont {{Szczepa{\'n}czyk}}, \citenamefont {{Gill}},\ and\
  \citenamefont {{Wongwathanarat}}}]{RiZaAn23}%
  \BibitemOpen
  \bibfield  {author} {\bibinfo {author} {\bibfnamefont {C.~J.}\ \bibnamefont
  {{Richardson}}}, \bibinfo {author} {\bibfnamefont {M.}~\bibnamefont
  {{Zanolin}}}, \bibinfo {author} {\bibfnamefont {H.}~\bibnamefont
  {{Andresen}}}, \bibinfo {author} {\bibfnamefont {M.~J.}\ \bibnamefont
  {{Szczepa{\'n}czyk}}}, \bibinfo {author} {\bibfnamefont {K.}~\bibnamefont
  {{Gill}}}, \ and\ \bibinfo {author} {\bibfnamefont {A.}~\bibnamefont
  {{Wongwathanarat}}},\ }\href {\doibase 10.1103/PhysRevD.105.103008}
  {\bibfield  {journal} {\bibinfo  {journal} {\prd}\ }\textbf {\bibinfo
  {volume} {105}},\ \bibinfo {eid} {103008} (\bibinfo {year} {2022})},\ \Eprint
  {http://arxiv.org/abs/2109.01582} {arXiv:2109.01582 [astro-ph.HE]}
  \BibitemShut {NoStop}%
\bibitem [{\citenamefont {Richers}\ \emph {et~al.}(2017)\citenamefont
  {Richers}, \citenamefont {Ott}, \citenamefont {Abdikamalov}, \citenamefont
  {O'Connor},\ and\ \citenamefont {Sullivan}}]{Richers_2017}%
  \BibitemOpen
  \bibfield  {author} {\bibinfo {author} {\bibfnamefont {S.}~\bibnamefont
  {Richers}}, \bibinfo {author} {\bibfnamefont {C.~D.}\ \bibnamefont {Ott}},
  \bibinfo {author} {\bibfnamefont {E.}~\bibnamefont {Abdikamalov}}, \bibinfo
  {author} {\bibfnamefont {E.}~\bibnamefont {O'Connor}}, \ and\ \bibinfo
  {author} {\bibfnamefont {C.}~\bibnamefont {Sullivan}},\ }\href {\doibase
  10.1103/physrevd.95.063019} {\bibfield  {journal} {\bibinfo  {journal}
  {Physical Review D}\ }\textbf {\bibinfo {volume} {95}} (\bibinfo {year}
  {2017}),\ 10.1103/physrevd.95.063019}\BibitemShut {NoStop}%
\bibitem [{\citenamefont {Scheidegger}\ \emph {et~al.}(2010)\citenamefont
  {Scheidegger}, \citenamefont {Käppeli}, \citenamefont {Whitehouse},
  \citenamefont {Fischer},\ and\ \citenamefont
  {Liebendörfer}}]{Scheidegger_2010}%
  \BibitemOpen
  \bibfield  {author} {\bibinfo {author} {\bibfnamefont {S.}~\bibnamefont
  {Scheidegger}}, \bibinfo {author} {\bibfnamefont {R.}~\bibnamefont
  {Käppeli}}, \bibinfo {author} {\bibfnamefont {S.~C.}\ \bibnamefont
  {Whitehouse}}, \bibinfo {author} {\bibfnamefont {T.}~\bibnamefont {Fischer}},
  \ and\ \bibinfo {author} {\bibfnamefont {M.}~\bibnamefont {Liebendörfer}},\
  }\href {\doibase 10.1051/0004-6361/200913220} {\bibfield  {journal} {\bibinfo
   {journal} {Astronomy and Astrophysics}\ }\textbf {\bibinfo {volume} {514}},\
  \bibinfo {pages} {A51} (\bibinfo {year} {2010})}\BibitemShut {NoStop}%
\bibitem [{\citenamefont {{Shibagaki}}\ \emph {et~al.}(2020)\citenamefont
  {{Shibagaki}}, \citenamefont {{Kuroda}}, \citenamefont {{Kotake}},\ and\
  \citenamefont {{Takiwaki}}}]{ShKuKo20}%
  \BibitemOpen
  \bibfield  {author} {\bibinfo {author} {\bibfnamefont {S.}~\bibnamefont
  {{Shibagaki}}}, \bibinfo {author} {\bibfnamefont {T.}~\bibnamefont
  {{Kuroda}}}, \bibinfo {author} {\bibfnamefont {K.}~\bibnamefont {{Kotake}}},
  \ and\ \bibinfo {author} {\bibfnamefont {T.}~\bibnamefont {{Takiwaki}}},\
  }\href {\doibase 10.1093/mnrasl/slaa021} {\bibfield  {journal} {\bibinfo
  {journal} {\mnras}\ }\textbf {\bibinfo {volume} {493}},\ \bibinfo {pages}
  {L138} (\bibinfo {year} {2020})},\ \Eprint {http://arxiv.org/abs/1909.09730}
  {arXiv:1909.09730 [astro-ph.HE]} \BibitemShut {NoStop}%
\bibitem [{\citenamefont {{Shibagaki}}\ \emph {et~al.}(2021)\citenamefont
  {{Shibagaki}}, \citenamefont {{Kuroda}}, \citenamefont {{Kotake}},\ and\
  \citenamefont {{Takiwaki}}}]{ShKuKo21}%
  \BibitemOpen
  \bibfield  {author} {\bibinfo {author} {\bibfnamefont {S.}~\bibnamefont
  {{Shibagaki}}}, \bibinfo {author} {\bibfnamefont {T.}~\bibnamefont
  {{Kuroda}}}, \bibinfo {author} {\bibfnamefont {K.}~\bibnamefont {{Kotake}}},
  \ and\ \bibinfo {author} {\bibfnamefont {T.}~\bibnamefont {{Takiwaki}}},\
  }\href {\doibase 10.1093/mnras/stab228} {\bibfield  {journal} {\bibinfo
  {journal} {\mnras}\ }\textbf {\bibinfo {volume} {502}},\ \bibinfo {pages}
  {3066} (\bibinfo {year} {2021})},\ \Eprint {http://arxiv.org/abs/2010.03882}
  {arXiv:2010.03882 [astro-ph.HE]} \BibitemShut {NoStop}%
\bibitem [{\citenamefont {{Srivastava}}\ \emph {et~al.}(2019)\citenamefont
  {{Srivastava}}, \citenamefont {{Ballmer}}, \citenamefont {{Brown}},
  \citenamefont {{Afle}}, \citenamefont {{Burrows}}, \citenamefont {{Radice}},\
  and\ \citenamefont {{Vartanyan}}}]{SrBaBr19}%
  \BibitemOpen
  \bibfield  {author} {\bibinfo {author} {\bibfnamefont {V.}~\bibnamefont
  {{Srivastava}}}, \bibinfo {author} {\bibfnamefont {S.}~\bibnamefont
  {{Ballmer}}}, \bibinfo {author} {\bibfnamefont {D.~A.}\ \bibnamefont
  {{Brown}}}, \bibinfo {author} {\bibfnamefont {C.}~\bibnamefont {{Afle}}},
  \bibinfo {author} {\bibfnamefont {A.}~\bibnamefont {{Burrows}}}, \bibinfo
  {author} {\bibfnamefont {D.}~\bibnamefont {{Radice}}}, \ and\ \bibinfo
  {author} {\bibfnamefont {D.}~\bibnamefont {{Vartanyan}}},\ }\href {\doibase
  10.1103/PhysRevD.100.043026} {\bibfield  {journal} {\bibinfo  {journal}
  {\prd}\ }\textbf {\bibinfo {volume} {100}},\ \bibinfo {eid} {043026}
  (\bibinfo {year} {2019})},\ \Eprint {http://arxiv.org/abs/1906.00084}
  {arXiv:1906.00084 [gr-qc]} \BibitemShut {NoStop}%
\bibitem [{\citenamefont {{Takiwaki}}\ and\ \citenamefont
  {{Kotake}}(2018)}]{TaKo18}%
  \BibitemOpen
  \bibfield  {author} {\bibinfo {author} {\bibfnamefont {T.}~\bibnamefont
  {{Takiwaki}}}\ and\ \bibinfo {author} {\bibfnamefont {K.}~\bibnamefont
  {{Kotake}}},\ }\href {\doibase 10.1093/mnrasl/sly008} {\bibfield  {journal}
  {\bibinfo  {journal} {\mnras}\ }\textbf {\bibinfo {volume} {475}},\ \bibinfo
  {pages} {L91} (\bibinfo {year} {2018})},\ \Eprint
  {http://arxiv.org/abs/1711.01905} {arXiv:1711.01905 [astro-ph.HE]}
  \BibitemShut {NoStop}%
\bibitem [{\citenamefont {{Vartanyan}}\ and\ \citenamefont
  {{Burrows}}(2020)}]{VaBu20}%
  \BibitemOpen
  \bibfield  {author} {\bibinfo {author} {\bibfnamefont {D.}~\bibnamefont
  {{Vartanyan}}}\ and\ \bibinfo {author} {\bibfnamefont {A.}~\bibnamefont
  {{Burrows}}},\ }\href {\doibase 10.3847/1538-4357/abafac} {\bibfield
  {journal} {\bibinfo  {journal} {\apj}\ }\textbf {\bibinfo {volume} {901}},\
  \bibinfo {eid} {108} (\bibinfo {year} {2020})},\ \Eprint
  {http://arxiv.org/abs/2007.07261} {arXiv:2007.07261 [astro-ph.HE]}
  \BibitemShut {NoStop}%
\bibitem [{\citenamefont {{Vartanyan}}\ \emph {et~al.}(2019)\citenamefont
  {{Vartanyan}}, \citenamefont {{Burrows}}, \citenamefont {{Radice}},
  \citenamefont {{Skinner}},\ and\ \citenamefont {{Dolence}}}]{VaBuRa19}%
  \BibitemOpen
  \bibfield  {author} {\bibinfo {author} {\bibfnamefont {D.}~\bibnamefont
  {{Vartanyan}}}, \bibinfo {author} {\bibfnamefont {A.}~\bibnamefont
  {{Burrows}}}, \bibinfo {author} {\bibfnamefont {D.}~\bibnamefont {{Radice}}},
  \bibinfo {author} {\bibfnamefont {M.~A.}\ \bibnamefont {{Skinner}}}, \ and\
  \bibinfo {author} {\bibfnamefont {J.}~\bibnamefont {{Dolence}}},\ }\href
  {\doibase 10.1093/mnras/sty2585} {\bibfield  {journal} {\bibinfo  {journal}
  {\mnras}\ }\textbf {\bibinfo {volume} {482}},\ \bibinfo {pages} {351}
  (\bibinfo {year} {2019})},\ \Eprint {http://arxiv.org/abs/1809.05106}
  {arXiv:1809.05106 [astro-ph.HE]} \BibitemShut {NoStop}%
\bibitem [{\citenamefont {Warren}\ \emph {et~al.}(2020)\citenamefont {Warren},
  \citenamefont {Couch}, \citenamefont {O'Connor},\ and\ \citenamefont
  {Morozova}}]{Warren_2020}%
  \BibitemOpen
  \bibfield  {author} {\bibinfo {author} {\bibfnamefont {M.~L.}\ \bibnamefont
  {Warren}}, \bibinfo {author} {\bibfnamefont {S.~M.}\ \bibnamefont {Couch}},
  \bibinfo {author} {\bibfnamefont {E.~P.}\ \bibnamefont {O'Connor}}, \ and\
  \bibinfo {author} {\bibfnamefont {V.}~\bibnamefont {Morozova}},\ }\href
  {\doibase 10.3847/1538-4357/ab97b7} {\bibfield  {journal} {\bibinfo
  {journal} {The Astrophysical Journal}\ }\textbf {\bibinfo {volume} {898}},\
  \bibinfo {pages} {139} (\bibinfo {year} {2020})}\BibitemShut {NoStop}%
\bibitem [{\citenamefont {{Yakunin}}\ \emph {et~al.}(2015)\citenamefont
  {{Yakunin}}, \citenamefont {{Mezzacappa}}, \citenamefont {{Marronetti}},
  \citenamefont {{Yoshida}}, \citenamefont {{Bruenn}}, \citenamefont {{Hix}},
  \citenamefont {{Lentz}}, \citenamefont {{Bronson Messer}}, \citenamefont
  {{Harris}}, \citenamefont {{Endeve}}, \citenamefont {{Blondin}},\ and\
  \citenamefont {{Lingerfelt}}}]{YaMeMa15}%
  \BibitemOpen
  \bibfield  {author} {\bibinfo {author} {\bibfnamefont {K.~N.}\ \bibnamefont
  {{Yakunin}}}, \bibinfo {author} {\bibfnamefont {A.}~\bibnamefont
  {{Mezzacappa}}}, \bibinfo {author} {\bibfnamefont {P.}~\bibnamefont
  {{Marronetti}}}, \bibinfo {author} {\bibfnamefont {S.}~\bibnamefont
  {{Yoshida}}}, \bibinfo {author} {\bibfnamefont {S.~W.}\ \bibnamefont
  {{Bruenn}}}, \bibinfo {author} {\bibfnamefont {W.~R.}\ \bibnamefont {{Hix}}},
  \bibinfo {author} {\bibfnamefont {E.~J.}\ \bibnamefont {{Lentz}}}, \bibinfo
  {author} {\bibfnamefont {O.~E.}\ \bibnamefont {{Bronson Messer}}}, \bibinfo
  {author} {\bibfnamefont {J.~A.}\ \bibnamefont {{Harris}}}, \bibinfo {author}
  {\bibfnamefont {E.}~\bibnamefont {{Endeve}}}, \bibinfo {author}
  {\bibfnamefont {J.~M.}\ \bibnamefont {{Blondin}}}, \ and\ \bibinfo {author}
  {\bibfnamefont {E.~J.}\ \bibnamefont {{Lingerfelt}}},\ }\href {\doibase
  10.1103/PhysRevD.92.084040} {\bibfield  {journal} {\bibinfo  {journal}
  {\prd}\ }\textbf {\bibinfo {volume} {92}},\ \bibinfo {eid} {084040} (\bibinfo
  {year} {2015})},\ \Eprint {http://arxiv.org/abs/1505.05824} {arXiv:1505.05824
  [astro-ph.HE]} \BibitemShut {NoStop}%
\bibitem [{\citenamefont {Astone}\ \emph {et~al.}(2018)\citenamefont {Astone},
  \citenamefont {Cerd{\'{a} }-Dur{\'{a}}n}, \citenamefont {Palma},
  \citenamefont {Drago}, \citenamefont {Muciaccia}, \citenamefont {Palomba},\
  and\ \citenamefont {Ricci}}]{Astone_2018}%
  \BibitemOpen
  \bibfield  {author} {\bibinfo {author} {\bibfnamefont {P.}~\bibnamefont
  {Astone}}, \bibinfo {author} {\bibfnamefont {P.}~\bibnamefont {Cerd{\'{a}
  }-Dur{\'{a}}n}}, \bibinfo {author} {\bibfnamefont {I.~D.}\ \bibnamefont
  {Palma}}, \bibinfo {author} {\bibfnamefont {M.}~\bibnamefont {Drago}},
  \bibinfo {author} {\bibfnamefont {F.}~\bibnamefont {Muciaccia}}, \bibinfo
  {author} {\bibfnamefont {C.}~\bibnamefont {Palomba}}, \ and\ \bibinfo
  {author} {\bibfnamefont {F.}~\bibnamefont {Ricci}},\ }\href {\doibase
  10.1103/physrevd.98.122002} {\bibfield  {journal} {\bibinfo  {journal}
  {Physical Review D}\ }\textbf {\bibinfo {volume} {98}} (\bibinfo {year}
  {2018}),\ 10.1103/physrevd.98.122002}\BibitemShut {NoStop}%
\bibitem [{\citenamefont {Bizouard}\ \emph {et~al.}(2021)\citenamefont
  {Bizouard}, \citenamefont {Maturana-Russel}, \citenamefont {Torres-Forn{\'{e}
  }}, \citenamefont {Obergaulinger}, \citenamefont {Cerd{\'{a}}-Dur{\'{a}}n},
  \citenamefont {Christensen}, \citenamefont {Font},\ and\ \citenamefont
  {Meyer}}]{Bizouard_2021}%
  \BibitemOpen
  \bibfield  {author} {\bibinfo {author} {\bibfnamefont {M.-A.}\ \bibnamefont
  {Bizouard}}, \bibinfo {author} {\bibfnamefont {P.}~\bibnamefont
  {Maturana-Russel}}, \bibinfo {author} {\bibfnamefont {A.}~\bibnamefont
  {Torres-Forn{\'{e} }}}, \bibinfo {author} {\bibfnamefont {M.}~\bibnamefont
  {Obergaulinger}}, \bibinfo {author} {\bibfnamefont {P.}~\bibnamefont
  {Cerd{\'{a}}-Dur{\'{a}}n}}, \bibinfo {author} {\bibfnamefont
  {N.}~\bibnamefont {Christensen}}, \bibinfo {author} {\bibfnamefont {J.~A.}\
  \bibnamefont {Font}}, \ and\ \bibinfo {author} {\bibfnamefont
  {R.}~\bibnamefont {Meyer}},\ }\href {\doibase 10.1103/physrevd.103.063006}
  {\bibfield  {journal} {\bibinfo  {journal} {Physical Review D}\ }\textbf
  {\bibinfo {volume} {103}} (\bibinfo {year} {2021}),\
  10.1103/physrevd.103.063006}\BibitemShut {NoStop}%
\bibitem [{\citenamefont {Torres-Forn{\'{e} }}\ \emph
  {et~al.}(2019)\citenamefont {Torres-Forn{\'{e} }}, \citenamefont
  {Cerd{\'{a}}-Dur{\'{a}}n}, \citenamefont {Obergaulinger}, \citenamefont
  {Müller},\ and\ \citenamefont {Font}}]{Torres_Forn__2019}%
  \BibitemOpen
  \bibfield  {author} {\bibinfo {author} {\bibfnamefont {A.}~\bibnamefont
  {Torres-Forn{\'{e} }}}, \bibinfo {author} {\bibfnamefont {P.}~\bibnamefont
  {Cerd{\'{a}}-Dur{\'{a}}n}}, \bibinfo {author} {\bibfnamefont
  {M.}~\bibnamefont {Obergaulinger}}, \bibinfo {author} {\bibfnamefont
  {B.}~\bibnamefont {Müller}}, \ and\ \bibinfo {author} {\bibfnamefont
  {J.~A.}\ \bibnamefont {Font}},\ }\href {\doibase
  10.1103/physrevlett.123.051102} {\bibfield  {journal} {\bibinfo  {journal}
  {Physical Review Letters}\ }\textbf {\bibinfo {volume} {123}} (\bibinfo
  {year} {2019}),\ 10.1103/physrevlett.123.051102}\BibitemShut {NoStop}%
\bibitem [{\citenamefont {Lin}\ \emph {et~al.}(2022)\citenamefont {Lin},
  \citenamefont {Rijal}, \citenamefont {Lunardini}, \citenamefont {Morales},\
  and\ \citenamefont {Zanolin}}]{https://doi.org/10.48550/arxiv.2211.07878}%
  \BibitemOpen
  \bibfield  {author} {\bibinfo {author} {\bibfnamefont {Z.}~\bibnamefont
  {Lin}}, \bibinfo {author} {\bibfnamefont {A.}~\bibnamefont {Rijal}}, \bibinfo
  {author} {\bibfnamefont {C.}~\bibnamefont {Lunardini}}, \bibinfo {author}
  {\bibfnamefont {M.~D.}\ \bibnamefont {Morales}}, \ and\ \bibinfo {author}
  {\bibfnamefont {M.}~\bibnamefont {Zanolin}},\ }\href {\doibase
  10.48550/ARXIV.2211.07878} {\enquote {\bibinfo {title} {Characterizing a
  supernova's standing accretion shock instability with neutrinos and
  gravitational waves},}\ } (\bibinfo {year} {2022})\BibitemShut {NoStop}%
\bibitem [{\citenamefont {Klimenko}\ \emph {et~al.}(2005)\citenamefont
  {Klimenko}, \citenamefont {Mohanty}, \citenamefont {Rakhmanov},\ and\
  \citenamefont {Mitselmakher}}]{Klimenko_2005}%
  \BibitemOpen
  \bibfield  {author} {\bibinfo {author} {\bibfnamefont {S.}~\bibnamefont
  {Klimenko}}, \bibinfo {author} {\bibfnamefont {S.}~\bibnamefont {Mohanty}},
  \bibinfo {author} {\bibfnamefont {M.}~\bibnamefont {Rakhmanov}}, \ and\
  \bibinfo {author} {\bibfnamefont {G.}~\bibnamefont {Mitselmakher}},\ }\href
  {\doibase 10.1103/physrevd.72.122002} {\bibfield  {journal} {\bibinfo
  {journal} {Physical Review D}\ }\textbf {\bibinfo {volume} {72}} (\bibinfo
  {year} {2005}),\ 10.1103/physrevd.72.122002}\BibitemShut {NoStop}%
\bibitem [{\citenamefont {Klimenko}\ \emph {et~al.}(2016)\citenamefont
  {Klimenko}, \citenamefont {Vedovato}, \citenamefont {Drago}, \citenamefont
  {Salemi}, \citenamefont {Tiwari}, \citenamefont {Prodi}, \citenamefont
  {Lazzaro}, \citenamefont {Ackley}, \citenamefont {Tiwari}, \citenamefont
  {Silva},\ and\ \citenamefont {Mitselmakher}}]{Klimenko_2016}%
  \BibitemOpen
  \bibfield  {author} {\bibinfo {author} {\bibfnamefont {S.}~\bibnamefont
  {Klimenko}}, \bibinfo {author} {\bibfnamefont {G.}~\bibnamefont {Vedovato}},
  \bibinfo {author} {\bibfnamefont {M.}~\bibnamefont {Drago}}, \bibinfo
  {author} {\bibfnamefont {F.}~\bibnamefont {Salemi}}, \bibinfo {author}
  {\bibfnamefont {V.}~\bibnamefont {Tiwari}}, \bibinfo {author} {\bibfnamefont
  {G.}~\bibnamefont {Prodi}}, \bibinfo {author} {\bibfnamefont
  {C.}~\bibnamefont {Lazzaro}}, \bibinfo {author} {\bibfnamefont
  {K.}~\bibnamefont {Ackley}}, \bibinfo {author} {\bibfnamefont
  {S.}~\bibnamefont {Tiwari}}, \bibinfo {author} {\bibfnamefont {C.~D.}\
  \bibnamefont {Silva}}, \ and\ \bibinfo {author} {\bibfnamefont
  {G.}~\bibnamefont {Mitselmakher}},\ }\href {\doibase
  10.1103/physrevd.93.042004} {\bibfield  {journal} {\bibinfo  {journal}
  {Physical Review D}\ }\textbf {\bibinfo {volume} {93}} (\bibinfo {year}
  {2016}),\ 10.1103/physrevd.93.042004}\BibitemShut {NoStop}%
\bibitem [{\citenamefont {Klimenko}\ \emph {et~al.}(2008)\citenamefont
  {Klimenko}, \citenamefont {Yakushin}, \citenamefont {Mercer},\ and\
  \citenamefont {Mitselmakher}}]{Klimenko_2008}%
  \BibitemOpen
  \bibfield  {author} {\bibinfo {author} {\bibfnamefont {S.}~\bibnamefont
  {Klimenko}}, \bibinfo {author} {\bibfnamefont {I.}~\bibnamefont {Yakushin}},
  \bibinfo {author} {\bibfnamefont {A.}~\bibnamefont {Mercer}}, \ and\ \bibinfo
  {author} {\bibfnamefont {G.}~\bibnamefont {Mitselmakher}},\ }\href {\doibase
  10.1088/0264-9381/25/11/114029} {\bibfield  {journal} {\bibinfo  {journal}
  {Classical and Quantum Gravity}\ }\textbf {\bibinfo {volume} {25}},\ \bibinfo
  {pages} {114029} (\bibinfo {year} {2008})}\BibitemShut {NoStop}%
\bibitem [{\citenamefont {Drago}\ \emph {et~al.}(2021)\citenamefont {Drago},
  \citenamefont {Klimenko}, \citenamefont {Lazzaro}, \citenamefont {Milotti},
  \citenamefont {Mitselmakher}, \citenamefont {Necula}, \citenamefont
  {O’Brian}, \citenamefont {Prodi}, \citenamefont {Salemi}, \citenamefont
  {Szczepanczyk}, \citenamefont {Tiwari}, \citenamefont {Tiwari}, \citenamefont
  {V}, \citenamefont {Vedovato},\ and\ \citenamefont
  {Yakushin}}]{DRAGO2021100678}%
  \BibitemOpen
  \bibfield  {author} {\bibinfo {author} {\bibfnamefont {M.}~\bibnamefont
  {Drago}}, \bibinfo {author} {\bibfnamefont {S.}~\bibnamefont {Klimenko}},
  \bibinfo {author} {\bibfnamefont {C.}~\bibnamefont {Lazzaro}}, \bibinfo
  {author} {\bibfnamefont {E.}~\bibnamefont {Milotti}}, \bibinfo {author}
  {\bibfnamefont {G.}~\bibnamefont {Mitselmakher}}, \bibinfo {author}
  {\bibfnamefont {V.}~\bibnamefont {Necula}}, \bibinfo {author} {\bibfnamefont
  {B.}~\bibnamefont {O’Brian}}, \bibinfo {author} {\bibfnamefont {G.~A.}\
  \bibnamefont {Prodi}}, \bibinfo {author} {\bibfnamefont {F.}~\bibnamefont
  {Salemi}}, \bibinfo {author} {\bibfnamefont {M.}~\bibnamefont
  {Szczepanczyk}}, \bibinfo {author} {\bibfnamefont {S.}~\bibnamefont
  {Tiwari}}, \bibinfo {author} {\bibfnamefont {V.}~\bibnamefont {Tiwari}},
  \bibinfo {author} {\bibfnamefont {G.}~\bibnamefont {V}}, \bibinfo {author}
  {\bibfnamefont {G.}~\bibnamefont {Vedovato}}, \ and\ \bibinfo {author}
  {\bibfnamefont {I.}~\bibnamefont {Yakushin}},\ }\href {\doibase
  https://doi.org/10.1016/j.softx.2021.100678} {\bibfield  {journal} {\bibinfo
  {journal} {SoftwareX}\ }\textbf {\bibinfo {volume} {14}},\ \bibinfo {pages}
  {100678} (\bibinfo {year} {2021})}\BibitemShut {NoStop}%
\bibitem [{\citenamefont {{Andresen}}\ \emph {et~al.}(2019)\citenamefont
  {{Andresen}}, \citenamefont {{M{\"u}ller}}, \citenamefont {{Janka}},
  \citenamefont {{Summa}}, \citenamefont {{Gill}},\ and\ \citenamefont
  {{Zanolin}}}]{2019MNRAS.486.2238A}%
  \BibitemOpen
  \bibfield  {author} {\bibinfo {author} {\bibfnamefont {H.}~\bibnamefont
  {{Andresen}}}, \bibinfo {author} {\bibfnamefont {E.}~\bibnamefont
  {{M{\"u}ller}}}, \bibinfo {author} {\bibfnamefont {H.~T.}\ \bibnamefont
  {{Janka}}}, \bibinfo {author} {\bibfnamefont {A.}~\bibnamefont {{Summa}}},
  \bibinfo {author} {\bibfnamefont {K.}~\bibnamefont {{Gill}}}, \ and\ \bibinfo
  {author} {\bibfnamefont {M.}~\bibnamefont {{Zanolin}}},\ }\href {\doibase
  10.1093/mnras/stz990} {\bibfield  {journal} {\bibinfo  {journal} {\mnras}\
  }\textbf {\bibinfo {volume} {486}},\ \bibinfo {pages} {2238} (\bibinfo {year}
  {2019})},\ \Eprint {http://arxiv.org/abs/1810.07638} {arXiv:1810.07638
  [astro-ph.HE]} \BibitemShut {NoStop}%
\bibitem [{\citenamefont {Szczepa{\'{n} }czyk}\ \emph
  {et~al.}(2021)\citenamefont {Szczepa{\'{n} }czyk}, \citenamefont {Antelis},
  \citenamefont {Benjamin}, \citenamefont {Cavagli{\`{a}}}, \citenamefont
  {Gondek-Rosi{\'{n}}ska}, \citenamefont {Hansen}, \citenamefont {Klimenko},
  \citenamefont {Morales}, \citenamefont {Moreno}, \citenamefont {Mukherjee},
  \citenamefont {Nurbek}, \citenamefont {Powell}, \citenamefont {Singh},
  \citenamefont {Sitmukhambetov}, \citenamefont {Szewczyk}, \citenamefont
  {Valdez}, \citenamefont {Vedovato}, \citenamefont {Westhouse}, \citenamefont
  {Zanolin},\ and\ \citenamefont {Zheng}}]{Szczepa_czyk_2021}%
  \BibitemOpen
  \bibfield  {author} {\bibinfo {author} {\bibfnamefont {M.~J.}\ \bibnamefont
  {Szczepa{\'{n} }czyk}}, \bibinfo {author} {\bibfnamefont {J.~M.}\
  \bibnamefont {Antelis}}, \bibinfo {author} {\bibfnamefont {M.}~\bibnamefont
  {Benjamin}}, \bibinfo {author} {\bibfnamefont {M.}~\bibnamefont
  {Cavagli{\`{a}}}}, \bibinfo {author} {\bibfnamefont {D.}~\bibnamefont
  {Gondek-Rosi{\'{n}}ska}}, \bibinfo {author} {\bibfnamefont {T.}~\bibnamefont
  {Hansen}}, \bibinfo {author} {\bibfnamefont {S.}~\bibnamefont {Klimenko}},
  \bibinfo {author} {\bibfnamefont {M.~D.}\ \bibnamefont {Morales}}, \bibinfo
  {author} {\bibfnamefont {C.}~\bibnamefont {Moreno}}, \bibinfo {author}
  {\bibfnamefont {S.}~\bibnamefont {Mukherjee}}, \bibinfo {author}
  {\bibfnamefont {G.}~\bibnamefont {Nurbek}}, \bibinfo {author} {\bibfnamefont
  {J.}~\bibnamefont {Powell}}, \bibinfo {author} {\bibfnamefont
  {N.}~\bibnamefont {Singh}}, \bibinfo {author} {\bibfnamefont
  {S.}~\bibnamefont {Sitmukhambetov}}, \bibinfo {author} {\bibfnamefont
  {P.}~\bibnamefont {Szewczyk}}, \bibinfo {author} {\bibfnamefont
  {O.}~\bibnamefont {Valdez}}, \bibinfo {author} {\bibfnamefont
  {G.}~\bibnamefont {Vedovato}}, \bibinfo {author} {\bibfnamefont
  {J.}~\bibnamefont {Westhouse}}, \bibinfo {author} {\bibfnamefont
  {M.}~\bibnamefont {Zanolin}}, \ and\ \bibinfo {author} {\bibfnamefont
  {Y.}~\bibnamefont {Zheng}},\ }\href {\doibase 10.1103/physrevd.104.102002}
  {\bibfield  {journal} {\bibinfo  {journal} {Physical Review D}\ }\textbf
  {\bibinfo {volume} {104}} (\bibinfo {year} {2021}),\
  10.1103/physrevd.104.102002}\BibitemShut {NoStop}%
\bibitem [{\citenamefont {Slutsky}\ \emph {et~al.}(2010)\citenamefont
  {Slutsky}, \citenamefont {Blackburn}, \citenamefont {Brown}, \citenamefont
  {Cadonati}, \citenamefont {Cain}, \citenamefont {Cavagli{\`{a} }},
  \citenamefont {Chatterji}, \citenamefont {Christensen}, \citenamefont
  {Coughlin}, \citenamefont {Desai}, \citenamefont {Gonz{\'{a}}lez},
  \citenamefont {Isogai}, \citenamefont {Katsavounidis}, \citenamefont
  {Rankins}, \citenamefont {Reed}, \citenamefont {Riles}, \citenamefont
  {Shawhan}, \citenamefont {Smith}, \citenamefont {Zotov},\ and\ \citenamefont
  {Zweizig}}]{Slutsky_2010}%
  \BibitemOpen
  \bibfield  {author} {\bibinfo {author} {\bibfnamefont {J.}~\bibnamefont
  {Slutsky}}, \bibinfo {author} {\bibfnamefont {L.}~\bibnamefont {Blackburn}},
  \bibinfo {author} {\bibfnamefont {D.~A.}\ \bibnamefont {Brown}}, \bibinfo
  {author} {\bibfnamefont {L.}~\bibnamefont {Cadonati}}, \bibinfo {author}
  {\bibfnamefont {J.}~\bibnamefont {Cain}}, \bibinfo {author} {\bibfnamefont
  {M.}~\bibnamefont {Cavagli{\`{a} }}}, \bibinfo {author} {\bibfnamefont
  {S.}~\bibnamefont {Chatterji}}, \bibinfo {author} {\bibfnamefont
  {N.}~\bibnamefont {Christensen}}, \bibinfo {author} {\bibfnamefont
  {M.}~\bibnamefont {Coughlin}}, \bibinfo {author} {\bibfnamefont
  {S.}~\bibnamefont {Desai}}, \bibinfo {author} {\bibfnamefont
  {G.}~\bibnamefont {Gonz{\'{a}}lez}}, \bibinfo {author} {\bibfnamefont
  {T.}~\bibnamefont {Isogai}}, \bibinfo {author} {\bibfnamefont
  {E.}~\bibnamefont {Katsavounidis}}, \bibinfo {author} {\bibfnamefont
  {B.}~\bibnamefont {Rankins}}, \bibinfo {author} {\bibfnamefont
  {T.}~\bibnamefont {Reed}}, \bibinfo {author} {\bibfnamefont {K.}~\bibnamefont
  {Riles}}, \bibinfo {author} {\bibfnamefont {P.}~\bibnamefont {Shawhan}},
  \bibinfo {author} {\bibfnamefont {J.~R.}\ \bibnamefont {Smith}}, \bibinfo
  {author} {\bibfnamefont {N.}~\bibnamefont {Zotov}}, \ and\ \bibinfo {author}
  {\bibfnamefont {J.}~\bibnamefont {Zweizig}},\ }\href {\doibase
  10.1088/0264-9381/27/16/165023} {\bibfield  {journal} {\bibinfo  {journal}
  {Classical and Quantum Gravity}\ }\textbf {\bibinfo {volume} {27}},\ \bibinfo
  {pages} {165023} (\bibinfo {year} {2010})}\BibitemShut {NoStop}%
\bibitem [{\citenamefont {Cavagli{\`{a} }}\ \emph {et~al.}(2020)\citenamefont
  {Cavagli{\`{a} }}, \citenamefont {Gaudio}, \citenamefont {Hansen},
  \citenamefont {Staats}, \citenamefont {Szczepa{\'{n}}czyk},\ and\
  \citenamefont {Zanolin}}]{Cavagli__2020}%
  \BibitemOpen
  \bibfield  {author} {\bibinfo {author} {\bibfnamefont {M.}~\bibnamefont
  {Cavagli{\`{a} }}}, \bibinfo {author} {\bibfnamefont {S.}~\bibnamefont
  {Gaudio}}, \bibinfo {author} {\bibfnamefont {T.}~\bibnamefont {Hansen}},
  \bibinfo {author} {\bibfnamefont {K.}~\bibnamefont {Staats}}, \bibinfo
  {author} {\bibfnamefont {M.}~\bibnamefont {Szczepa{\'{n}}czyk}}, \ and\
  \bibinfo {author} {\bibfnamefont {M.}~\bibnamefont {Zanolin}},\ }\href
  {\doibase 10.1088/2632-2153/ab527d} {\bibfield  {journal} {\bibinfo
  {journal} {Machine Learning: Science and Technology}\ }\textbf {\bibinfo
  {volume} {1}},\ \bibinfo {pages} {015005} (\bibinfo {year}
  {2020})}\BibitemShut {NoStop}%
\bibitem [{\citenamefont {Necula}\ \emph {et~al.}(2012)\citenamefont {Necula},
  \citenamefont {Klimenko},\ and\ \citenamefont {Mitselmakher}}]{Necula_2012}%
  \BibitemOpen
  \bibfield  {author} {\bibinfo {author} {\bibfnamefont {V.}~\bibnamefont
  {Necula}}, \bibinfo {author} {\bibfnamefont {S.}~\bibnamefont {Klimenko}}, \
  and\ \bibinfo {author} {\bibfnamefont {G.}~\bibnamefont {Mitselmakher}},\
  }\href {\doibase 10.1088/1742-6596/363/1/012032} {\bibfield  {journal}
  {\bibinfo  {journal} {Journal of Physics: Conference Series}\ }\textbf
  {\bibinfo {volume} {363}},\ \bibinfo {pages} {012032} (\bibinfo {year}
  {2012})}\BibitemShut {NoStop}%
\bibitem [{\citenamefont {Cuoco}\ \emph {et~al.}(2020)\citenamefont {Cuoco},
  \citenamefont {Powell}, \citenamefont {Cavagli{\`{a} }}, \citenamefont
  {Ackley}, \citenamefont {Bejger}, \citenamefont {Chatterjee}, \citenamefont
  {Coughlin}, \citenamefont {Coughlin}, \citenamefont {Easter}, \citenamefont
  {Essick}, \citenamefont {Gabbard}, \citenamefont {Gebhard}, \citenamefont
  {Ghosh}, \citenamefont {Haegel}, \citenamefont {Iess}, \citenamefont
  {Keitel}, \citenamefont {M{\'{a}}rka}, \citenamefont {M{\'{a}}rka},
  \citenamefont {Morawski}, \citenamefont {Nguyen}, \citenamefont {Ormiston},
  \citenamefont {Pürrer}, \citenamefont {Razzano}, \citenamefont {Staats},
  \citenamefont {Vajente},\ and\ \citenamefont {Williams}}]{Cuoco_2020}%
  \BibitemOpen
  \bibfield  {author} {\bibinfo {author} {\bibfnamefont {E.}~\bibnamefont
  {Cuoco}}, \bibinfo {author} {\bibfnamefont {J.}~\bibnamefont {Powell}},
  \bibinfo {author} {\bibfnamefont {M.}~\bibnamefont {Cavagli{\`{a} }}},
  \bibinfo {author} {\bibfnamefont {K.}~\bibnamefont {Ackley}}, \bibinfo
  {author} {\bibfnamefont {M.}~\bibnamefont {Bejger}}, \bibinfo {author}
  {\bibfnamefont {C.}~\bibnamefont {Chatterjee}}, \bibinfo {author}
  {\bibfnamefont {M.}~\bibnamefont {Coughlin}}, \bibinfo {author}
  {\bibfnamefont {S.}~\bibnamefont {Coughlin}}, \bibinfo {author}
  {\bibfnamefont {P.}~\bibnamefont {Easter}}, \bibinfo {author} {\bibfnamefont
  {R.}~\bibnamefont {Essick}}, \bibinfo {author} {\bibfnamefont
  {H.}~\bibnamefont {Gabbard}}, \bibinfo {author} {\bibfnamefont
  {T.}~\bibnamefont {Gebhard}}, \bibinfo {author} {\bibfnamefont
  {S.}~\bibnamefont {Ghosh}}, \bibinfo {author} {\bibfnamefont
  {L.}~\bibnamefont {Haegel}}, \bibinfo {author} {\bibfnamefont
  {A.}~\bibnamefont {Iess}}, \bibinfo {author} {\bibfnamefont {D.}~\bibnamefont
  {Keitel}}, \bibinfo {author} {\bibfnamefont {Z.}~\bibnamefont {M{\'{a}}rka}},
  \bibinfo {author} {\bibfnamefont {S.}~\bibnamefont {M{\'{a}}rka}}, \bibinfo
  {author} {\bibfnamefont {F.}~\bibnamefont {Morawski}}, \bibinfo {author}
  {\bibfnamefont {T.}~\bibnamefont {Nguyen}}, \bibinfo {author} {\bibfnamefont
  {R.}~\bibnamefont {Ormiston}}, \bibinfo {author} {\bibfnamefont
  {M.}~\bibnamefont {Pürrer}}, \bibinfo {author} {\bibfnamefont
  {M.}~\bibnamefont {Razzano}}, \bibinfo {author} {\bibfnamefont
  {K.}~\bibnamefont {Staats}}, \bibinfo {author} {\bibfnamefont
  {G.}~\bibnamefont {Vajente}}, \ and\ \bibinfo {author} {\bibfnamefont
  {D.}~\bibnamefont {Williams}},\ }\href {\doibase 10.1088/2632-2153/abb93a}
  {\bibfield  {journal} {\bibinfo  {journal} {Machine Learning: Science and
  Technology}\ }\textbf {\bibinfo {volume} {2}},\ \bibinfo {pages} {011002}
  (\bibinfo {year} {2020})}\BibitemShut {NoStop}%
\bibitem [{\citenamefont {George}\ and\ \citenamefont
  {Huerta}(2018)}]{George_2018}%
  \BibitemOpen
  \bibfield  {author} {\bibinfo {author} {\bibfnamefont {D.}~\bibnamefont
  {George}}\ and\ \bibinfo {author} {\bibfnamefont {E.}~\bibnamefont
  {Huerta}},\ }\href {\doibase 10.1103/physrevd.97.044039} {\bibfield
  {journal} {\bibinfo  {journal} {Physical Review D}\ }\textbf {\bibinfo
  {volume} {97}} (\bibinfo {year} {2018}),\
  10.1103/physrevd.97.044039}\BibitemShut {NoStop}%
\bibitem [{\citenamefont {Antelis}\ \emph {et~al.}(2022)\citenamefont
  {Antelis}, \citenamefont {Cavaglia}, \citenamefont {Hansen}, \citenamefont
  {Morales}, \citenamefont {Moreno}, \citenamefont {Mukherjee}, \citenamefont
  {Szczepa{\'{n} }czyk},\ and\ \citenamefont {Zanolin}}]{Antelis_2022}%
  \BibitemOpen
  \bibfield  {author} {\bibinfo {author} {\bibfnamefont {J.~M.}\ \bibnamefont
  {Antelis}}, \bibinfo {author} {\bibfnamefont {M.}~\bibnamefont {Cavaglia}},
  \bibinfo {author} {\bibfnamefont {T.}~\bibnamefont {Hansen}}, \bibinfo
  {author} {\bibfnamefont {M.~D.}\ \bibnamefont {Morales}}, \bibinfo {author}
  {\bibfnamefont {C.}~\bibnamefont {Moreno}}, \bibinfo {author} {\bibfnamefont
  {S.}~\bibnamefont {Mukherjee}}, \bibinfo {author} {\bibfnamefont {M.~J.}\
  \bibnamefont {Szczepa{\'{n} }czyk}}, \ and\ \bibinfo {author} {\bibfnamefont
  {M.}~\bibnamefont {Zanolin}},\ }\href {\doibase 10.1103/physrevd.105.084054}
  {\bibfield  {journal} {\bibinfo  {journal} {Physical Review D}\ }\textbf
  {\bibinfo {volume} {105}} (\bibinfo {year} {2022}),\
  10.1103/physrevd.105.084054}\BibitemShut {NoStop}%
\bibitem [{\citenamefont {Chan}\ \emph {et~al.}(2020)\citenamefont {Chan},
  \citenamefont {Heng},\ and\ \citenamefont {Messenger}}]{Chan_2020}%
  \BibitemOpen
  \bibfield  {author} {\bibinfo {author} {\bibfnamefont {M.~L.}\ \bibnamefont
  {Chan}}, \bibinfo {author} {\bibfnamefont {I.~S.}\ \bibnamefont {Heng}}, \
  and\ \bibinfo {author} {\bibfnamefont {C.}~\bibnamefont {Messenger}},\ }\href
  {\doibase 10.1103/physrevd.102.043022} {\bibfield  {journal} {\bibinfo
  {journal} {Physical Review D}\ }\textbf {\bibinfo {volume} {102}} (\bibinfo
  {year} {2020}),\ 10.1103/physrevd.102.043022}\BibitemShut {NoStop}%
\bibitem [{\citenamefont {{Mukherjee}}\ \emph {et~al.}(2017)\citenamefont
  {{Mukherjee}}, \citenamefont {{Salazar}}, \citenamefont {{Mittelstaedt}},\
  and\ \citenamefont {{Valdez}}}]{Mukherjee:2017}%
  \BibitemOpen
  \bibfield  {author} {\bibinfo {author} {\bibfnamefont {S.}~\bibnamefont
  {{Mukherjee}}}, \bibinfo {author} {\bibfnamefont {L.}~\bibnamefont
  {{Salazar}}}, \bibinfo {author} {\bibfnamefont {J.}~\bibnamefont
  {{Mittelstaedt}}}, \ and\ \bibinfo {author} {\bibfnamefont {O.}~\bibnamefont
  {{Valdez}}},\ }\href {\doibase 10.1103/PhysRevD.96.104033} {\bibfield
  {journal} {\bibinfo  {journal} {\prd}\ }\textbf {\bibinfo {volume} {96}},\
  \bibinfo {eid} {104033} (\bibinfo {year} {2017})}\BibitemShut {NoStop}%
\bibitem [{\citenamefont {L{\'{o} }pez}\ \emph {et~al.}(2021)\citenamefont
  {L{\'{o} }pez}, \citenamefont {Palma}, \citenamefont {Drago}, \citenamefont
  {Cerd{\'{a}}-Dur{\'{a}}n},\ and\ \citenamefont {Ricci}}]{L_pez_2021}%
  \BibitemOpen
  \bibfield  {author} {\bibinfo {author} {\bibfnamefont {M.}~\bibnamefont
  {L{\'{o} }pez}}, \bibinfo {author} {\bibfnamefont {I.~D.}\ \bibnamefont
  {Palma}}, \bibinfo {author} {\bibfnamefont {M.}~\bibnamefont {Drago}},
  \bibinfo {author} {\bibfnamefont {P.}~\bibnamefont
  {Cerd{\'{a}}-Dur{\'{a}}n}}, \ and\ \bibinfo {author} {\bibfnamefont
  {F.}~\bibnamefont {Ricci}},\ }\href {\doibase 10.1103/physrevd.103.063011}
  {\bibfield  {journal} {\bibinfo  {journal} {Physical Review D}\ }\textbf
  {\bibinfo {volume} {103}} (\bibinfo {year} {2021}),\
  10.1103/physrevd.103.063011}\BibitemShut {NoStop}%
\bibitem [{\citenamefont {Morales}\ \emph {et~al.}(2021)\citenamefont
  {Morales}, \citenamefont {Antelis}, \citenamefont {Moreno},\ and\
  \citenamefont {Nesterov}}]{Morales2020}%
  \BibitemOpen
  \bibfield  {author} {\bibinfo {author} {\bibfnamefont {M.~D.}\ \bibnamefont
  {Morales}}, \bibinfo {author} {\bibfnamefont {J.~M.}\ \bibnamefont
  {Antelis}}, \bibinfo {author} {\bibfnamefont {C.}~\bibnamefont {Moreno}}, \
  and\ \bibinfo {author} {\bibfnamefont {A.~I.}\ \bibnamefont {Nesterov}},\
  }\href {\doibase 10.3390/s21093174} {\bibfield  {journal} {\bibinfo
  {journal} {Sensors}\ }\textbf {\bibinfo {volume} {21}},\ \bibinfo {pages}
  {3175} (\bibinfo {year} {2021})}\BibitemShut {NoStop}%
\bibitem [{\citenamefont {Bishop}(2006)}]{Bishop2006}%
  \BibitemOpen
  \bibfield  {author} {\bibinfo {author} {\bibfnamefont {C.~M.}\ \bibnamefont
  {Bishop}},\ }\href@noop {} {\emph {\bibinfo {title} {Pattern Recognition and
  Machine Learning (Information Science and Statistics)}}}\ (\bibinfo
  {publisher} {Springer-Verlag New York, Inc.},\ \bibinfo {address} {Secaucus,
  NJ, USA},\ \bibinfo {year} {2006})\BibitemShut {NoStop}%
\bibitem [{\citenamefont {Burrows}\ and\ \citenamefont
  {Vartanyan}(2021)}]{Burrows_2021}%
  \BibitemOpen
  \bibfield  {author} {\bibinfo {author} {\bibfnamefont {A.}~\bibnamefont
  {Burrows}}\ and\ \bibinfo {author} {\bibfnamefont {D.}~\bibnamefont
  {Vartanyan}},\ }\href {\doibase 10.1038/s41586-020-03059-w} {\bibfield
  {journal} {\bibinfo  {journal} {Nature}\ }\textbf {\bibinfo {volume} {589}},\
  \bibinfo {pages} {29} (\bibinfo {year} {2021})}\BibitemShut {NoStop}%
\bibitem [{\citenamefont {{Woosley}}\ and\ \citenamefont
  {{Weaver}}(1995)}]{WoWe95}%
  \BibitemOpen
  \bibfield  {author} {\bibinfo {author} {\bibfnamefont {S.~E.}\ \bibnamefont
  {{Woosley}}}\ and\ \bibinfo {author} {\bibfnamefont {T.~A.}\ \bibnamefont
  {{Weaver}}},\ }\href {\doibase 10.1086/192237} {\bibfield  {journal}
  {\bibinfo  {journal} {\apjs}\ }\textbf {\bibinfo {volume} {101}},\ \bibinfo
  {pages} {181} (\bibinfo {year} {1995})}\BibitemShut {NoStop}%
\bibitem [{\citenamefont {{Woosley}}\ and\ \citenamefont
  {{Heger}}(2007)}]{WoHe07}%
  \BibitemOpen
  \bibfield  {author} {\bibinfo {author} {\bibfnamefont {S.~E.}\ \bibnamefont
  {{Woosley}}}\ and\ \bibinfo {author} {\bibfnamefont {A.}~\bibnamefont
  {{Heger}}},\ }\href {\doibase 10.1016/j.physrep.2007.02.009} {\bibfield
  {journal} {\bibinfo  {journal} {\physrep}\ }\textbf {\bibinfo {volume}
  {442}},\ \bibinfo {pages} {269} (\bibinfo {year} {2007})},\ \Eprint
  {http://arxiv.org/abs/astro-ph/0702176} {arXiv:astro-ph/0702176 [astro-ph]}
  \BibitemShut {NoStop}%
\end{thebibliography}%

\end{document}